\documentclass[11pt,a4paper]{article}
\usepackage{jcappub}
\pdfoutput=1
\usepackage{graphicx}
\usepackage{longtable}
\usepackage{color}

\title{Big Bounce Genesis and Possible Experimental Tests -- A Brief Review} 

\author[1]{Yeuk-Kwan~E.~Cheung,}

\author[2]{and Changhong~Li, } 

\author[1, 3, 4,5]{J.D. Vergados }

\affiliation[1]{Department of Physics, Nanjing University, 22 Hankou Road, Nanjing, China 210093;} 

\affiliation[2]{Department of Astronomy, Key Laboratory of Astroparticle Physics of Yunnan Province, Yunnan University, No.2 Cuihu North Road, Kunming, China 650091; }
\affiliation[3]{ ARC Centre of Excellence in Particle Physics at the Terascale and Centre for the Subatomic Structure of Matter (CSSM), University of Adelaide, Adelaide SA 5005, Australia; }
\affiliation[4]{TEI of Western Macedonia, Kozani, Gr 501 00,  Greece}
\affiliation[5]{Permanent address: Theoretical Physics,University of Ioannina, Ioannina, Gr 451 10, Greece.} 

\emailAdd{cheung@nju.edu.cn} 
\emailAdd{changhongli@ynu.edu.cn} 
\emailAdd{vergados@uoi.gr} 

\abstract{We review the recent status of big bounce genesis 
as a new possibility of using dark matter particle's mass and  interaction cross section to test the existence of 
a  bounce universe  at the early stage of evolution in our  currently observed universe.  
To study the dark matter production and evolution inside the 
bounce universe, called big bounce genesis for short, we propose 
a model independent approach.  
 We shall present the motivation for proposing  big bounce as well the   model independent predictions  which can be tested by dark matter direct searches.  
 A positive finding shall  have profound impact on our understanding 
 of the early universe physics.
} 
\arxivnumber{0000.0000} 
\begin{document} 
\maketitle

\section{Motivation and Overview}
\label{sec:motivation}

Understanding the working principles of the fundamental constituents pivots  on the knowledge of the origin 
of universe~\cite{Laozi}. Cosmology is therefore the oldest 
intellectual  pursuit of mankind. 
The  inflationary paradigm is the laurel of modern cosmology
as it solves in one stroke {\it  the Monopoles, Horizon, } and {\it Flatness} problems of  the Big Bang Cosmology~(BBC) by modeling a brief period of  exponential expansion after the Big Bang  using simple scalar fields~\cite{Guth:1980zm}. 
It turns out that a nearly scale
 invariant primordial power spectrum, which agrees well with the current array of   Cosmic Microwave Background~(CMB) 
 observations~\cite{Komatsu:2010fb, Ade:2013kta},  
 can be generated from the quantum fluctuations of the scalar
 field   in the simplest inflation model~\cite{Mukhanov:1990me}. 
 This achievement is  crowned the ``Inflation paradigm.''


Slowly it dawns on some  cosmologists  that, 
similar to the BBC,
 the scenario of inflation also suffers from its own   problems, 
 the {\it initial singularity problem} and the  {\it fine-tuning problem}~\cite{Borde:1993xh}. 

Challenge presents great opportunities in developing new theories for the early universe. 
In past  decade in  a  concord effort  to resolve the initial singularity problem of the inflation scenario,  the ``bounce universe scenario''~(BUS), was proposed 
by postulating a  phase of contraction before the universe
turns around--called the Big Bounce  in lieu of the Big Bang
~(Recent reviews can be found~\cite{NoBer08, Branden12, 
Battefeld:2014uga,  Brandenberger:2016vhg}.).

\begin{figure}[htbp]
\begin{center}
\includegraphics[width=1.0\textwidth]{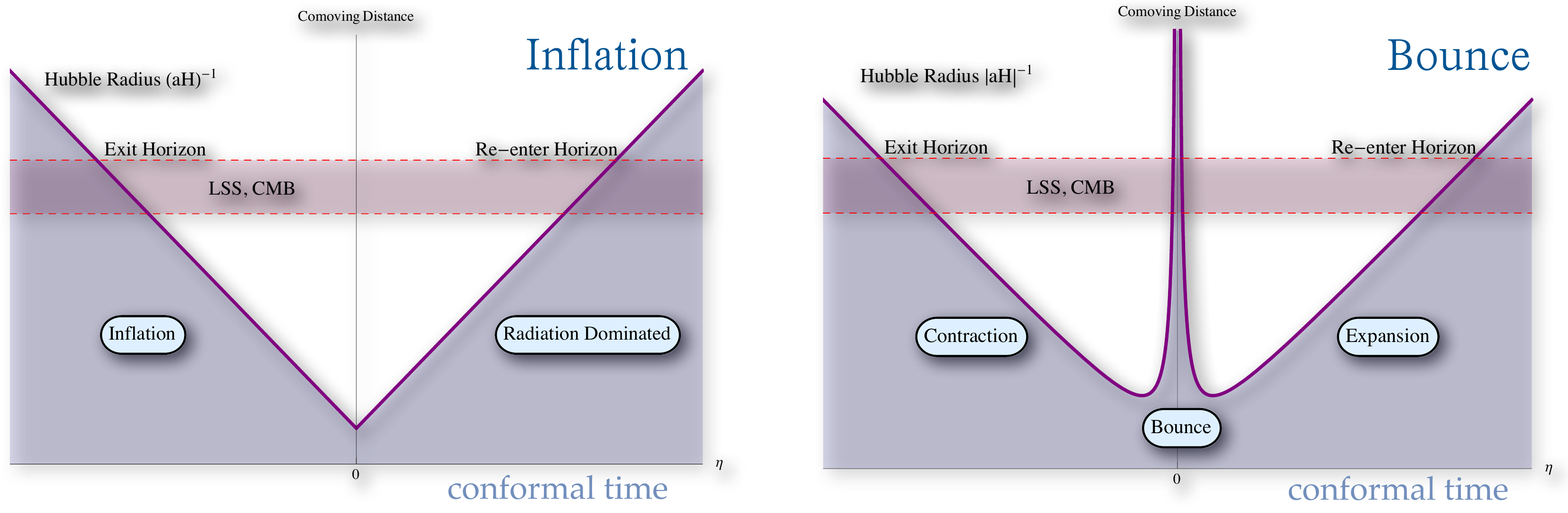}
\caption{The exit and entrance of physical perturbations at the Hubble Horizon in the Inflation~(left) and in the bounce~(right) scenarios.}
\label{horizons}
\end{center}
\end{figure}

Effort on detailed implementations ensue with many working 
models of the Big Bounce universe  proposed~\cite{Khoury:2001wf, 
Steinhardt:2001st, Steinhardt:2002ih, Gasperini:2002bn, Creminelli:2006xe, Cai:2007qw, Cai:2008qw, Wands:2008tv, Bhattacharya:2013ut, Odintsov:2014gea, Li:2014msi, 
Quintin:2014oea, Wan:2014fra, Cai:2014bea, Liu:2014tda, 
Li:2014qwa, Cai:2014hja, Cai:2014xxa, Hu:2014aua, Li:2014cka, Xia:2014tda, Nojiri:2015sfd, Nojiri:2016ygo, 
Cheung:2016oab, Escofet:2015gpa, Haro:2015oqa, 
Odintsov:2016tar, Choudhury:2015baa, Odintsov:2015ynk, 
Oikonomou:2015qha, Cai:2014jla, Cai:2015vzv, 
Ferreira:2016gfg, Quintin:2015rta, Brandenberger:2016egn, Hipolito-Ricaldi:2016kqq, Wan:2015hya, Cai:2014zga}.  
According to the  BUS, our universe 
is bouncing from a contracting phase to an expanding phase at non-zero minimum size so the Big Bang Singularity 
is resolved. 
The  Horizon Problem  and Flatness problems  are solved  by observing  that there is an  interplay of the 
physical scale and the Hubble scale similar to the inflationary 
scenario, as depicted in Fig.~\ref{horizons}. 
The solution to the Big Bang Problems--Horizon, Flatness and Homogeneity--by the CSTB model is explicitly demonstrated in a recent paper~\cite{Cheung:2016oab}.

Moreover, a {\it{stable}} as well as scale-invariant power spectrum of primordial density perturbations 
 matching up to the  currently observed 
 CMB spectra  can  be obtained  during the phase of matter 
dominated contraction in the BUS~\cite{Wands:1998yp, Finelli:2001sr, Li:2013bha, Li:2011nj}.
 
In light of such fast developments, we are well motivated to
 work out further criteria for testing the bounce universe 
 models,  and  extract discriminating predictions to 
 distinguish  the bounce scenario from the  inflationary paradigm. 
 Even though the details of cosmic evolution in the inflation scenario and BUS are so different, the experimental or observational evidence, which can be used to tell these two models apart, is still lacking.  
One may expect that the precisely measured CMB spectra are suitable for distinguish these two scenario. 
However, they are still not enough to concretely distinguish these two scenarios 
due to following two factors~\cite{Li:2014cba}:
\begin{enumerate}
\item In terms of primordial power spectrum, a complete duality between the inflation scenario and BUS have been well 
established~\cite{Wands:1998yp, Finelli:2001sr, Boyle:2004gv,
 Li:2012vi,  Li:2013bha}. 
It enables both inflation and BUS to generate
 a scale-invariant primordial power spectrum 
 with the same probability from 
  the unified parameter space. 
 In short, if a primordial power spectrum can be generated in an expanding phase of cosmological evolution, it
  can also  be generated in a contracting phase with the same 
  scale-dependence and time-dependence~\cite{Li:2013bha}. 
 Literally, they are degenerate in the leading-order 
 signatures of CMB spectra;\\
\item Currently, all models in these two scenarios are utilizing some undetected classical/quantum fields to drive the inflation or big bounce at the very  early stage of cosmic evolution. 
Hence, their predictions of the CMB spectrum  
and the scalar-tensor ratio, both built upon the linear perturbation theory of these unconfirmed fields, are still questionable.
\end{enumerate}
Therefore, the usual temperature-temperature correlations and scalar to tensor ratio in the  CMB spectrum  cannot serve as direct evidence for either the inflation scenario or BUS. 
Hence  new concrete methods--independent of CMB observations--for distinguishing the inflation scenario and BUS  with falsifiable predictions  are  urgently needed. 
That is where our proposal~\cite{Li:2014era}
 dubbed ``Big Bounce Genesis'' steps in, 
 in which dark matter(DM) direct detections experiments 
are proposed to be a testing ground to distinguish Bounce 
Universe Scenario from the inflation paradigm.

\begin{figure}[!ht]
\begin{center}
\includegraphics[scale=0.5]{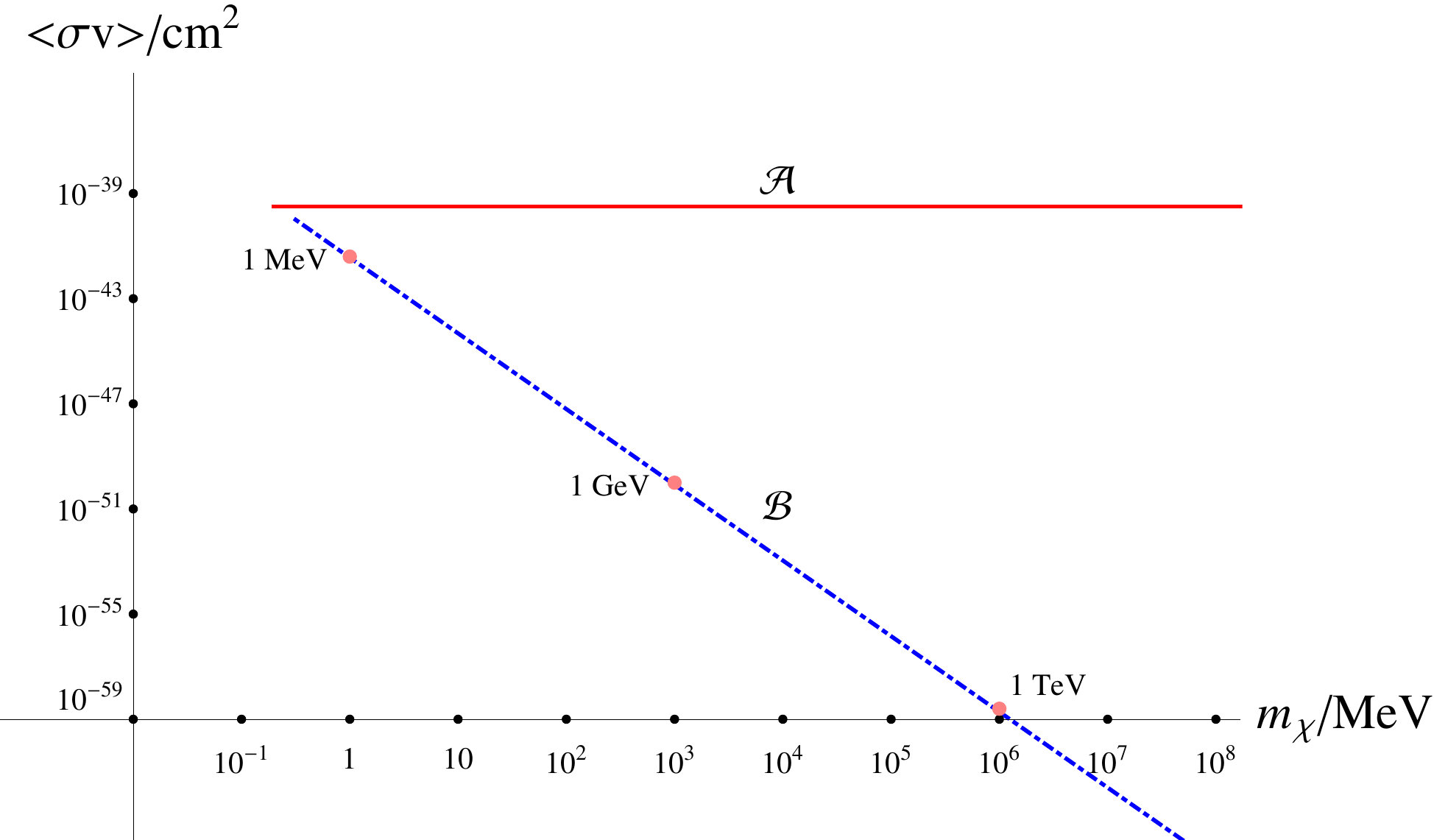}\\
\caption{The cross section $\langle\sigma \upsilon\rangle$ as a function of the scalar DM mass. In the standard cosmology it is a constant (solid line), but it varies considerably in the bounce universe scenario (dash-dotted line)~\cite{Li:2014era}.}
\label{fig:cbplog}
\end{center} 
\end{figure}

Big Bounce Genesis~(BBG)\cite{Li:2014era} 
is a framework for analyzing matter production and evolution in 
the bounce universe, in which there is a period of contraction 
prior to the a period of expansion connected by a bounce. 
By  incorporating the concept of 
{\it weak freeze-out}, an  explicit computation in 
an out-of-chemical equilibrium productions of DM 
 shows that 
the cross-section and DM mass is constrained by 
the observed relic abundance of DM, $\Omega_\chi$,~\cite{Li:2014era}:
\begin{equation}
\Omega_\chi\propto \langle\sigma v\rangle_\chi m_\chi^2~.
\end{equation}
This relation is depicted by Curve~B in  Fig.~\ref{fig:cbplog}.
This characteristic relation is significantly different from
 the Standard Cosmology predictions:
 \begin{equation}
  \Omega_\chi\propto\langle\sigma v\rangle_\chi^{-1} m_\chi^0, 
 \end{equation}
in either  the  WIMP or  WIMP-less 
  miracles~\cite{Scherrer:1985zt, Feng:2008ya, Kolb:1990vq,Gondolo:1990dk}, as depicted by Curve~A in  Fig.~\ref{fig:cbplog}

The existence of this relation  can therefore 
be a telling sign that the 
universe has gone through a Big Bounce~\cite{Li:2014era}.

Therefore one can be  hopeful that 
data from current and future DM 
detections can tell these two early universe scenarios 
apart~\cite{Cheung:2014pea}. 
A confirmed relation between DM mass and interaction cross section  not only can lend support to the idea that
the cosmos  goes through a bounce and not a bang, 
but also establishes the  one-way production of  matter 
 in the early universe. 
All of these shall  have profound implications 
in early universe physics.

In BBG~\cite{Li:2014era, Li:2014cba, Cheung:2014nxi}, DM particles are assumed to be produced by the annihilation
 of SM particles in hot plasma of the bounce  cosmological
 background. The   following model-independent interaction
 is assumed, 
\begin{equation}\label{eq:phitchi}
\phi+\phi\Longleftrightarrow \chi+\chi
\end{equation}
where $\phi$ denotes SM particles and $\chi$ denotes DM particles (can be either fermions or bosons). This assumption leaves DM mass, $m_\chi$, and cross-section, $\langle\sigma v\rangle$, as two free parameters constrained by the
 astrophysical observations. 
  
\begin{figure}[!ht]
\begin{center}
\includegraphics[scale=0.45]{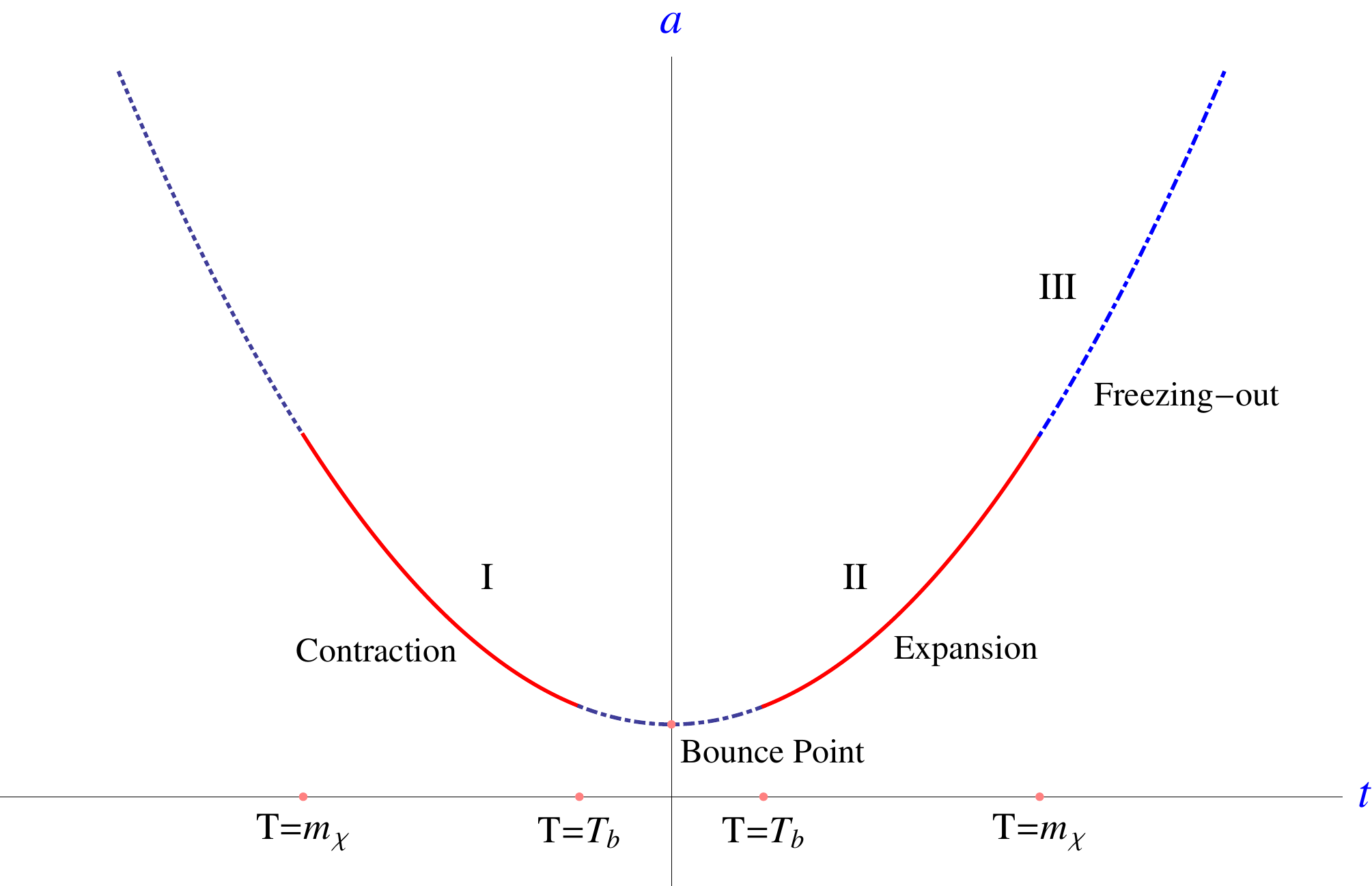}\\
 \caption{The breakdown of the Big bounce period into a pre-bounce contraction (phase I), a post-bounce expansion (phase II), and  the  freeze-out  of the DM particles (phase III).}
 \label{fig:bounce}
 \end{center}
  \end{figure}

 In a model independent analysis of the bounce universe 
 dynamics we divide  the  cosmic evolution of a generic bounce universe  schematically into three stages, 
   Phase I: {\it pre-bounce contraction},  
   Phase II: {\it post-bounce expansion} and 
   Phase III: {\it freeze-out phase} 
   as shown in Fig.~\ref{fig:bounce}.  
 
  If DM is produced efficiently in both of the pre-bounce contraction and post-bounce expansion when 
   the background temperature is high and the  duration is long enough. The  freeze-out process of DM commences after the post-expansion. Generically, such a thermal production mechanism is model independent and irrelevant to the details of 
   realization of bounce since the bounce point 
   connecting Phase I and Phase II is  assumed too short 
   to affect  DM productions~\cite{Cai:2011ci, Li:2014cba}. 
  
For the production of DM, there are two different avenues, {\it thermal equilibrium production} and {\it Out-of-chemical equilibrium production}.
\begin{itemize}
\item {\it Thermal equilibrium production:} DM particle with large cross-section are produced very efficiently, so that its abundance increases rapidly and achieves its thermal equilibrium value even in the pre-bounce contracting or post-bounce expanding phases. Then the abundance of DM tracks the thermal equilibrium value before the freeze-out takes place, 
\begin{equation}\label{eq:chep}
Y(t)=Y_{eq},\quad t<t_f~,
\end{equation}
where $Y\equiv \frac{n_\chi}{T^3}$ is the abundance of DM and $Y_{eq}$ is the thermal equilibrium abundance in the given cosmological background with temperature $T$, and $t_f$ is the moment of the freeze-out commencing. \\

\item  {\it Out-of-chemical equilibrium production}~\cite{Li:2014era}{\it :} In a given cosmological background, if the cross-section of DM is  small enough, the production of DM should be inefficient. Therefore, its abundance cannot achieve the thermal equilibrium value during the production process,
\begin{equation}\label{eq:ochep}
Y(t)\ll Y_{eq},\quad t<t_f~.
\end{equation}

\end{itemize}

And for the freeze-out process, there is also two different ways. At the end of Phase II: post-bounce expansion, the background temperature continues to fall as long as universe is expanding, the forward reaction of Eq.(\ref{eq:phitchi}) is suppressed exponentially, {\it i.e.} the production of DM is end. Depending on the abundance of DM at this moment, this freeze-out process of DM can be categorized into two types: {\it Strong freeze-out } and {\it Weak freeze-out}~\cite{Li:2014era}.
\begin{itemize}
\item {\it Strong freeze-out:} If a plenty of DM particles have been produced before, the backward reaction of Eq.(\ref{eq:phitchi}) becomes dominated as the forward reaction of Eq.(\ref{eq:phitchi}) is suppressed exponentially. The backward reaction decreases the abundance of DM very efficiently until the number density of DM is too low to keep thermal contact in the expanding phase. Therefore, after such strong freeze-out, the relic abundance of DM is significantly lower than that before freeze-out and is inverse to the DM cross-section,
\begin{equation}\label{eq:sfo}
Y_f\propto \frac{1}{\langle\sigma v\rangle},
\end{equation}
where $Y_f$ is the relic abundance of DM after freeze-out, $ t\ge t_f$. This is just the well-known freeze-out process in WIMP and WIMP-less miracle~\cite{Scherrer:1985zt, Feng:2008ya}, and we label it as ``strong freeze-out'' comparing with the ``weak freeze-out'' at following.  \\

\item {\it Weak freeze-out:} If the abundance of DM is very low, the backward reaction in Eq.(\ref{eq:phitchi}) is always negligible. When the forward reaction in Eq.(\ref{eq:phitchi}) is suppressed exponentially, both the production and annihilation of DM are end. Therefore, the relic abundance of DM is equal to the abundance of DM at the end of the production phases and is generically proportional to the DM cross-section,
\begin{equation}\label{eq:wfo}
Y_f=Y(t_f)\propto \langle\sigma v\rangle.
\end{equation} 
Remarkably, since all abundance of DM, which is sensitive to cosmological evolution, are preserved, such relic abundance of DM undergoing the {\it weak freeze-out} process is encoded with information of early evolution of universe. 
\end{itemize}

 In view of the  two possible  production  routes 
   and the subsequent two possible freeze-out processes of DM, 
   there arise four possibilities, 
as displayed in Table~\ref{tab:rout}.    
\begin{table*}[ht!]  
\vspace{-0.5cm}
\caption{\label{tab:rout}Production and freeze-out of DM in BUS}
\begin{center}
\begin{tabular}{|c|c|c|}
\hline
& Thermal equilibrium production & Out-of-chemical equilibrium production \\
\hline
Strong freeze-out& {\bf Route I} & ---\\
\hline
Weak freeze-out & {\bf Route III} & {\bf Route II}\\
\hline
\end{tabular}
\end{center}
\vspace{-0.5cm}
\end{table*}

\begin{figure}[!ht]
\begin{center}
\includegraphics[scale=0.55]{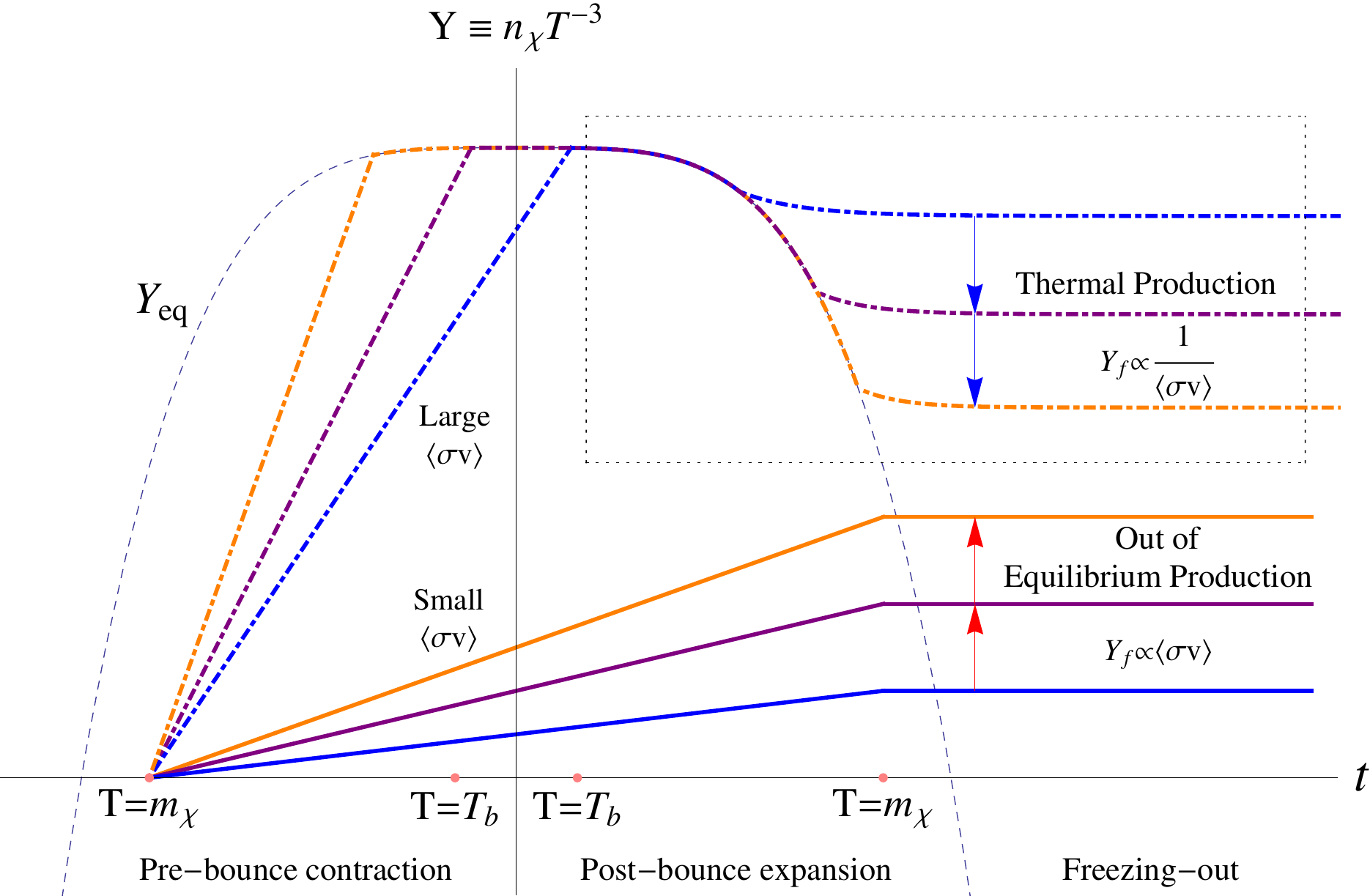}
 \caption{A schematic plot of the time evolution of DM in a generic bounce universe scenario. Two pathways of producing DM yet satisfying current observations thermal production (which is indistinguishable from standard cosmology) and non-thermal production (characteristic to bounce universe)  are illustrated. The horizontal axis indicates both the time, t, as well as the temperature, T, of the cosmological background~\cite{Li:2014era}.}
 \label{fig:RelicEvolution}
 \end{center}
 \end{figure}

The time evolution of DM in a generic bounce cosmos
 is illustrated in Fig.~\ref{fig:RelicEvolution}, 
 respectively, for {\bf Route I} and {\bf Route II}. 
 In following discussion, we take DM as a bosonic 
 particle and the highest temperature of bounce larger than the mass of DM (In this case, the {\bf Route III} is not manifest and we are discussing the details of the {\bf Route III} in next section.). 
  
In {\bf Route I} of BBG, DM is produced through {\it Thermal equilibrium production} and undergoes {\it Strong freeze-out}. DM commences its thermal decoupling from the thermal equilibrium state and, eventually, freezes out with the relic abundance inverse to its cross-section. Therefore, the relic abundance of DM predicted by {\bf Route I} is identical to the prediction of the models in the standard inflationary cosmology such as WIMP and WIMP-less miracles~\cite{Scherrer:1985zt, Feng:2008ya,  Li:2014era}. Particularly, the WIMP and WIMP-less miracles can be viewed as a part of this process, see the dashed black frame in Fig.~\ref{fig:RelicEvolution}. The relation of cross-section and mass predicted by this case is the Branch A in Fig.~\ref{fig:cbplog}  .

 
The novelty appears in the {\bf Route II}, {\it Out-of-chemical equilibrium production} and {\it Weak freeze-out}. During the production phase, the abundance of DM is much lower than the value of the thermal equilibrium state. With the falling of the temperature of background, DM takes a very weak freeze-out process that all pieces of information of early universe evolution are preserved in the relic abundance of DM, which leads a new characteristic relation of DM cross-section and mass that compatible with current observations({\it c.f.} Branch B in Table II of~\cite{Li:2014era}),
\begin{equation}\label{eq:macsbbg}
\langle\sigma v\rangle=7.2\times 10^{-26}m_\chi^{-2}~,\quad m_\chi\gg 432~\text{eV}
\end{equation}  
where $\langle\sigma v\rangle$ is the thermally averaged cross-section of DM. This relation of cross-section and mass is shown as the Branch B in Fig.~\ref{fig:cbplog}. As a smoking gun signal for the existence of the bounce universe, this novel relation can be used to check against recent and near future data from experiments of DM to determine whether or not universe undergoes a big bounce at a very early stage of cosmic evolution~\cite{Cheung:2014pea, Vergados:2016niz}.

In general  the production and evolution of DM in  the bounce cosmos (Big Bounce Genesis for short) brings about new possibilities, compared to the Standard model particle physics in the standard cosmology,  which are listed in Table~\ref{tab:cato}.
The generic picture of BBG is illustrated with the model of the evolution of scalar DM in a high temperature bounce. 
\begin{table*}[ht!]  
\vspace{-0.5cm}
\caption{\label{tab:cato}Categories of BBG models}
\begin{center}
\begin{tabular}{|c|c|c|}
\hline
& High Temperature Bounce & Low Temperature Bounce \\
& $T_b\gg m_\chi$ & $T_b\ll m_\chi$\\
\hline
Bosonic Dark Matter& {\bf Type I} & {\bf Type III}\\
\hline
Fermonic Dark Matter & {\bf Type II} & {\bf Type IV}\\
\hline
\end{tabular}
\end{center}
\vspace{-0.5cm}
\end{table*}
\\
In each of the venues  listed in Table~\ref{tab:cato}, 
DM is produced and evolves through different routes listed in Table~\ref{tab:rout} and gives (beyond standard model) predictions. 
In this review, we  discuss each case in detail.

This review is organized as follows: 
In section~\ref{sec:2},  we discuss each type of BBG dynamics  following~\cite{Li:2014era, Li:2014cba, Cheung:2014nxi}.  We  study the generation of thermal fluctuations  of DM in the bounce cosmos following~\cite{Li:2015egy}, and show that the bounce dynamic is stable against thermal fluctuations for $T_{bounce}\le T_{Planck}$, in section \ref{sec:3}.
In section \ref{sec:4}, we discuss the search of BBG DM in direct detection experiments for mass range DM mass$\sim 100 GeV$ using nuclear recoil~\cite{Cheung:2014pea} and for light DM by electron recoil~\cite{Vergados:2016niz}.

\section{Dark matter Production and Evolution in the Bounce Universe Scenario}
\label{sec:2}
In general, the  evolution of DM in BUS is governed by the Boltzmann equation, 
\begin{equation} \label{eq:boe}
\frac{d(n_\chi a^3)}{a^3dt}=\widetilde{\langle\sigma v\rangle}\left[\left(n_\chi^{eq}\right)^2-n_\chi^2\right]~, 
\end{equation}
where $n_\chi^{eq}$ is the thermal equilibrium number density of DM, $a$,  the scale factor of the cosmological background, and, $\widetilde{\langle\sigma v\rangle}$, the thermally averaged cross section with temperature dependence. Since the DM mass is large, the production phases of DM is radiation-dominated, 
\begin{equation}~\label{eq:rdr}
\rho\propto a^{-4}\propto T^4,
\end{equation} 
where $\rho$ is the energy density of cosmological background. 
To facilitate the study of whole evolution of DM in a generic bounce comos, without loss of generality we take following two conditions,
\begin{itemize}
\item Initial abundance of DM takes $n_\chi^{i}=0$, {\it i.e.} the number density of DM is set to be zero at the onset of pre-bounce contraction phase in which $T\ll m_\chi$.
\item Matching condition on bounce point is $n_\chi^+(T_b)=n_\chi^-(T_b)$, {\it i.e.}  the number density of DM, ${n_\chi}$, at the end of the pre-bounce contraction (denoted by $-$)  is equal to the initial abundance of the post-bounce expansion (denoted by $+$), given that the entropy of universe is conserved around the bounce point~\cite{Cai:2011ci}.
\end{itemize}
Therefore, by solving Eq.~\ref{eq:boe} with these two conditions, the evolution of DM abundance are fully determined. Then we can obtain the characteristic relations of DM cross section and mass for each type of BBG models constrained by recent observational energy fraction of DM, $\Omega_\chi=0.26$.

\subsection{Type I: Scalar Dark Matter in a High Temperature Bounce}
Following~\cite{Li:2014era}, the simplest case is that the highest temperature of bounce is larger than DM mass, $T_b\gg m_\chi$, and both $\chi$ and $\phi$ are scalar particles. Thus the interactions of DM with the light boson can be modeled by $\mathcal{L}_{int} = \lambda \phi^{2} \chi^{2}$. And, in the limit $m_\phi\rightarrow 0$, we get~\cite{Peskin:1995ev},
\begin{equation} \label{eq:std}
\widetilde{\langle\sigma v\rangle}=\left\{  
\begin{array} {l}
 {\displaystyle \frac{x^2}{4}\cdot\langle\sigma v\rangle ~, \qquad m_\chi\ll T}  \\ 
 \\ 
  {\displaystyle  \langle\sigma v\rangle ~, \qquad\qquad m_\chi\gg T}    \\
\end{array}     
\right. , \qquad x\equiv\frac{m_\chi}{T},\qquad \langle\sigma v\rangle= \frac{1}{32\pi} \frac{\lambda^2}{m_\chi^2}~.
\end{equation}
Substituting Eq.~\ref{eq:std} and Eq.~\ref{eq:rdr} into Eq.~\ref{eq:boe}, the Boltzmann equation can be simplified during the production phases,
\begin{equation}\label{eq:yev}
\frac{dY_\mp}{dx}=\mp f \langle\sigma v\rangle m_\chi(1-\pi^4Y_\mp^2)~,\quad x< 1,
\end{equation}
where $\mp$ denotes the pre-bounce contraction and post-bounce expansion respectively, $T\gg m_\chi$~, and $f$ is a constant during these phases with $f\equiv \frac{m_\chi^2}{4\pi^2} (|H|x^2)^{-1}=1.5\times 10^{26}~eV$, being constrained by recent astrophysical observations. Then, by solving Eq.~\ref{eq:yev} with the initial abundance of DM and matching condition on bounce point, the complete solution of the DM abundance in the post-bounce expansion is obtained, 
\begin{equation} 
\label{eq:yp}
Y_+=\frac{1-e^{-2\pi^2f\langle\sigma v\rangle m_\chi(1+x-x_b)}}{\pi^2\left(1+e^{-2\pi^2f\langle\sigma v\rangle m_\chi(1+x-2x_b)}\right)}~. 
\end{equation} 
At the end of DM production, $T\sim m_\chi$, this solution can be categorized into two limits, {\it Thermal equilibrium production} and {\it Out-of-chemical equilibrium production},
\begin{equation}  
\label{eq:ype}
Y_+|_{x=1}=
\left\{  
  \begin{array} {lr}
 {\displaystyle \pi^{-2}, \qquad\qquad~  4\pi^2f\langle\sigma v\rangle m_\chi\gg 1} 
 \\ 
  {\displaystyle 2f\langle\sigma v\rangle m_\chi~, \quad~ 4\pi^2f\langle\sigma v\rangle m_\chi \ll 1}   
  \\
\end{array}     
\right. .
\end{equation}
They are the two  venues of DM production discussed in last section. 
\begin{itemize}
\item {\it Thermal equilibrium production:} For the upper line in Eq.~\ref{eq:ype}, with the large value of $\langle\sigma v\rangle m_\chi$, DM is produced in plenty abundance which have reached the thermal equilibrium before the end of the production phases. 
\item {\it Out-of-chemical equilibrium production:} And for the lower line in Eq.~\ref{eq:ype} with the small value of $\langle\sigma v\rangle m_\chi$, the production is mostly oneway and thermal equilibrium cannot be established, so that its abundance is much lower than the value of thermal equilibrium state even at the end of he production phases.
\end{itemize}

After the production phases in which $x\le 1$, the cosmos is still in expansion and  the background temperature of  universe continues to fall. The production of DM is exponentially suppressed and thermal decoupling commence.  To determine the relic abundance of DM after freeze-out, we are solving the Boltzmann equation Eq.~\ref{eq:boe} in the low temperature region,
\begin{equation}  \label{eq:yil}
\frac{d Y}{d x}=4f\langle\sigma v\rangle m_\chi\left(\frac{\pi}{8}xe^{-2x}-\pi^4\frac{Y^2}{x^2}\right)~, \quad x\ge 1
\end{equation}
where the first term on the right hand side of Eq.(\ref{eq:yil}) is subdominant and hence discarded for $x>1$. 
Integrating it from $x=1$ to $x\rightarrow\infty$, we obtain the relic abundance of DM after freeze-out,
\begin{equation}  \label{eq:dYdxsol}
Y_f\equiv Y|_{x\rightarrow\infty}=\frac{1}{4\pi^4f\langle\sigma v\rangle m_\chi +(Y_+|_{x=1})^{-1}}~,
\end{equation}
which leads two distinctive outcomes of the freeze-out process:

\begin{itemize}

\item {\it Strong freeze-out:}  If $Y_+|_{x=1}\gg (4\pi^4f\langle\sigma v\rangle m_\chi)^{-1}$, the initial abundance of DM at the onset of the freeze-out process is large enough for pair-annihilation of DM particles during the thermal decoupling, so that the relic abundance of DM becomes irrelevant of the initial abundance. Particularly, it is inversely proportional to the cross section,  
\begin{equation}
Y_f= \frac{1.71 \times 10^{-29} eV^{-1}}{\langle\sigma v\rangle m_\chi}~.
\end{equation}

\item {\it Weak freeze-out:} If, on the other hand, $Y_+|_{x=1}\ll(4\pi^4f\langle\sigma v\rangle m_\chi)^{-1}$ and the density of DM is too low to pair-annihilate during the thermal decoupling. The relic abundance of DM after freeze-out in this limit is just the initial abundance at the onset of the freeze-out process,  
\begin{equation}
Y_f= Y_+|_{x=1}~.
\end{equation}

\end{itemize}
Notice that two criteria for production from Eq.~\ref{eq:ype} and for freeze-out process from Eq.~\ref{eq:dYdxsol} are mostly coincident. Therefore, the {\bf Route III}, {\it thermal equilibrium production} and {\it weak freeze-out}, is not manifest in {\bf Type I} model of BBG. In this model, if DM is produced in thermal equilibrium, it must undergo strong 
freeze-out; and if DM is produced through the out-of-chemical equilibrium production, it must undergo weak freeze-out, 
{\it i.e.}, only {\bf Route I} and {\bf Route II} are viable. 

For {\bf Route I},  as the abundance of DM tracks the thermal 
equilibrium  values during the production phase and 
the freeze-out phase, all information of the early universe 
encoded in the relic abundance of DM is washed out. 
This is analogous to the standard cosmology in which DM freeze out strongly from  equilibrium~\cite{Kolb:1990vq}, in both 
the  WIMP miracle scenario~\cite{Scherrer:1985zt} and the 
WIMP-less miracle scenario~\cite{Feng:2008ya}. 
Therefore the standard model and the bounce universe 
cannot be distinguished from each other as long as DM is 
produced at thermal equilibrium, depicted in 
Branch A of Fig.~\ref{fig:cbplog}.

In the production  {\bf Route II}, the abundance of DM has 
not reached thermal equilibrium and therefore not suppressed by 
the ``thermal envelop'' and its final abundance is directly 
proportional to its interaction cross sections: 
\begin{equation}
Y_f= Y_+|_{x=1}=3\times 10^{26}{\rm eV} \langle\sigma v\rangle m_\chi~,
\end{equation} 
where we have used Eq.~\ref{eq:ype}.  
The relic abundance of DM produced this way encodes the information of the early universe dynamics when we reconstruct its decoupling time from its interaction cross section. 
Utilizing the current observed value of $\Omega_\chi$,
\begin{equation}
\Omega_\chi=1.18\times 10^{-2}{\rm eV}\times m_\chi Y_f=0.26~,
\end{equation}
the DM relic abundance is  constrained. This in turn 
forces the DM  mass and cross section  to lie on the   characteristic relation,  Eq.~\ref{eq:macsbbg}, 
depicted as the Branch B in Fig.~\ref{fig:cbplog}. 

If the DM mass and its interaction cross section is found 
to be related, as one point in Branch B, by 
  (in)-direct detection of DM, we can confidently 
  conclude that the matter is produced  in the early universe 
   out of thermal equilibrium. 
   This is in fact shared by many a non-standard universe models 
 based on $f(R)$ gravity~\cite{Capozziello:2007ec, Capozziello:2012ie}.

\subsection{Type III and IV: Bosonic and Fermionic Dark Matter in a Low Temperature Bounce}
In a low temperature bounce, $T_b\ll m_\chi$, the DM is expected to be produced inefficiently. So we focus on\footnote{On the other hand, if the factor $f \langle\sigma v\rangle_0 m_\chi$ is very large, and $x_b\equiv \frac{m_\chi}{T_b}\rightarrow \mathcal{O}(1)$, the production of DM would be very efficient, so that the abundance of DM is able to achieve its thermal equilibrium, for this case, see~\cite{Li:2014era} . }  the route in which DM is produced in out-of-chemical equilibrium and undergoes weak freeze-out, {\it i.e.} {\bf Route II} listed in Table~\ref{tab:rout}, following~\cite{Li:2014era, Cheung:2014nxi}.  

During a low temperature bounce, $x_b\gg 1$, the thermally averaged cross section of DM, $\widetilde{\langle\sigma v\rangle}$,  becomes irrelevant of temperature at the leading order, 
\begin{equation}
\widetilde{\langle\sigma v\rangle}=\langle\sigma v\rangle_0+\mathcal{O}(x^{-1}),
\end{equation} 
where $\langle\sigma v\rangle_0$ is simply determined by the interacting coupling constant of DM particle~\cite{Gondolo:1990dk}. At here we approximate $\widetilde{\langle\sigma v\rangle}=\langle\sigma v\rangle_0$ for simplicity.  By utilizing the dimensionless abundance of DM, $\tilde{Y}\equiv n_\chi s^{-1}$, with the entropy density, $s$, the Boltzmann equation can be simplified in further as following,
\begin{equation}\label{eq:eomy}
\frac{d\tilde{Y}_{\mp}}{dx}=-\frac{s}{xH}\langle\sigma v\rangle_0\left(\tilde{Y}^2_\mp-\tilde{Y}_{eq}^2\right)
\end{equation}
where the subscript $\mp$ corresponds the sign of Hubble parameter, {\it i.e.} the contracting phase and expanding phase of a generic bounce universe respectively. The entropy density is given by $s=(2\pi^2/45)h_\star T^3$ with $h_\star$ being the relativistic degree of freedom for the entropy density. In the radiation dominated era, the Hubble parameter is proportional to $T^2$, $H=\frac{\pi T^2}{M_p}\sqrt{\frac{g_\star}{90}}$, where $g_\star$ and $M_p$ are, respectively, the relativistic degree of freedom of energy density and the reduce Planck mass.

In low temperature limit $x\gg 1$, the thermal equilibrium number density of DM is suppressed exponentially by $x$, $n_{eq}=g_\chi\left(m_\chi^2/2\pi x\right)^{\frac{3}{2}}e^{-x}$, with $g_\chi$ being the number of degree of freedom of $\chi$.  Thus the particle creation term, $\tilde{Y}_{eq}^2$, on the right-hand of Eq.(\ref{eq:eomy}) is suppressed exponentially by $2x$, $\tilde{Y}_{eq}^2\propto e^{-2x}$. Therefore, the production of DM can be expected to be very inefficient during such low temperature bounce, so that the abundance of DM are unlikely able to achieve its thermal equilibrium value. Under this consideration, we  focus on this out of thermal equilibrium case for the DM production, $\tilde{Y}\ll \tilde{Y}_{eq}$,  {\it i.e.} the annihilation term $\tilde{Y}_\mp^2$ on the right-hand of Eq.(\ref{eq:eomy}) can be dropped out in the following calculations.

By integrating Eq.(\ref{eq:eomy}) with the initial condition $\tilde{Y}_{-}^i=0$ and the match condition $\tilde{Y}_-(x_b)=\tilde{Y}_+(x_b)$, we obtain the relic abundance of DM, (c.f. Eq (3.3)in~\cite{Cheung:2014nxi}, which was firstly obtained in~\cite{Li:2014era})
\begin{equation}
\tilde{Y}_f=\tilde{Y}^+(x\gg x_b)=\mathcal{C}\langle\sigma v\rangle_0 m_\chi e^{-2x_b}(1+2x_b)~.
\end{equation}
where $\mathcal{C}=0.014M_p g_\star^{-\frac{1}{2}} h_\star^{-1}g_\chi^2$. By taking $g_\chi=1$ and $h_\star\simeq g_\star=90$ during the phases under consideration, $\mathcal{C}=6.9\times 10^{22} $eV. Imposing the currently observed value of $\Omega_\chi$~,
\begin{equation}\label{eq:ocyt}
\Omega_\chi=5.7 \times 10^8  m_\chi \tilde{Y}_f\text{GeV}^{-1}=0.26~,
\end{equation}
leads a precisely observational constrains on $\langle\sigma v\rangle_0$, $m_\chi$ and $x_b$,
\begin{equation}\label{eq:smmtb}
\langle\sigma v\rangle_0 m_\chi^2 e^{-2x_b}(1+2x_b)=6.6\times 10^{-24}~.
\end{equation}
To sum up, by utilizing this observational constrains, the highest temperature of bounce can be determined in this scenario with the given value of  $(\langle\sigma v\rangle_0, m_\chi)$.

\subsection{Type II: Fermionic Dark Matter in a High Temperature Bounce}
Following~\cite{Cheung:2014nxi}, in high temperature bounce $T_b \gg m_\chi$, the thermally averaged cross section of DM, $\widetilde{\langle\sigma v\rangle}$, can be parameterized as following,
\begin{equation}\label{eq:csfh}
\widetilde{\langle\sigma v\rangle}=\tilde{\sigma}_0  x^{-n}~,\quad x<1
\end{equation}
where $n>0$ and typically, $n=2$ for the pair annihilation processes of Dirac and Majorana fermions into a pair of massless fermions, and $\tilde{\sigma}_0$ is simply determined by the interacting coupling constant of DM particle and independent of temperature.

\begin{figure}[htp] 
\centering
\includegraphics[width=0.5\textwidth]{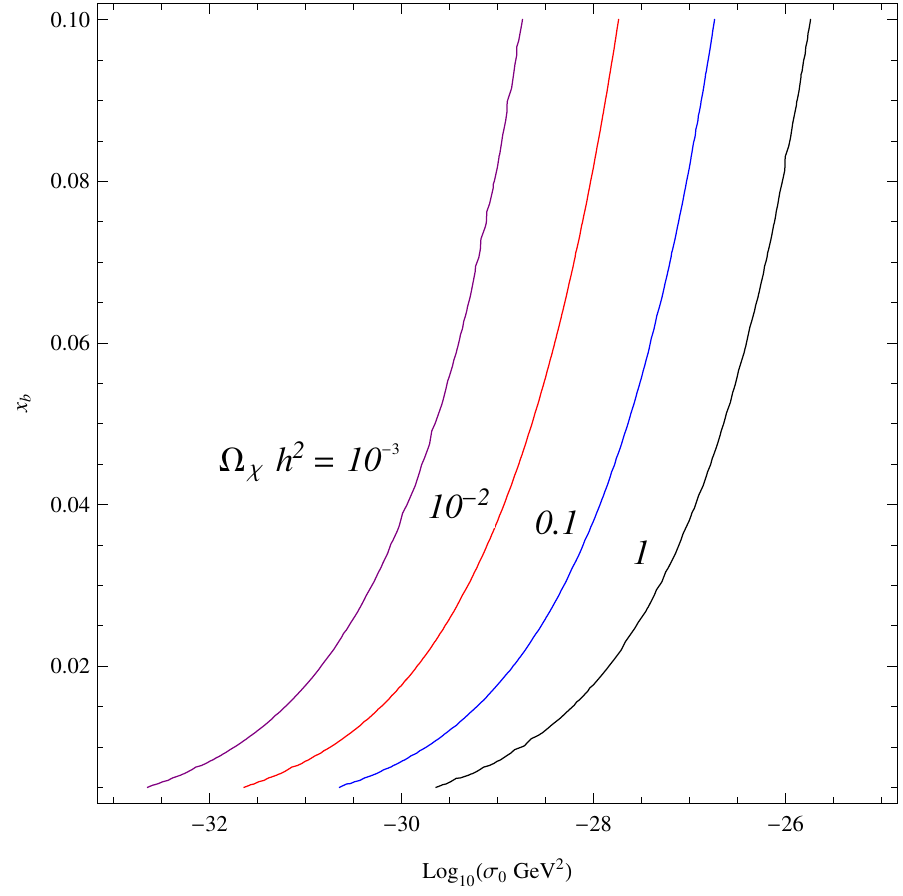}
\caption{Contour plots of the predicted relic abundance in the ($x_b$-$\tilde{\sigma}_0$) plane in high temperature bounce case. Here we take $m_\chi=1{\rm GeV}$.~\cite{Cheung:2014nxi}}\label{fig:fig_high_a-x-Omega-eps-converted-to.pdf}
\end{figure}

\begin{figure}[htp] 
\centering
\includegraphics[width=0.5\textwidth]{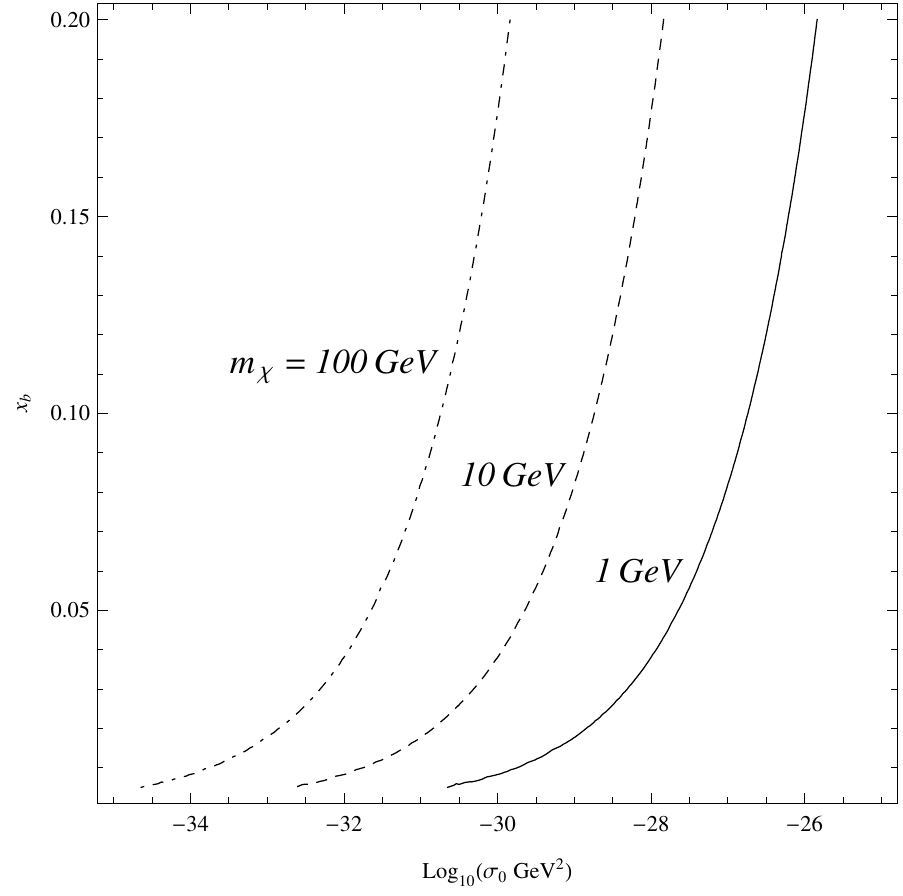}
\caption{Low temperature bounce: $x_b$ as a function of  $a$ to reproduce the present relic abundance $\Omega_\chi h^2\simeq0.1$ for various $m_\chi$.~\cite{Cheung:2014nxi}}\label{fig:fig_high_a-x-m-eps-converted-to.pdf}
\end{figure}

We again  focus on an interesting route in  which DM is 
produced out of chemical equilibrium and undergoes weak 
freeze-out. Then, by substituting Eq.~\ref{eq:csfh} into the Boltzmann equation, Eq.~\ref{eq:boe}, and integrating 
it with the two matching conditions, we get a solution
\begin{equation}
\tilde{Y}_{\pm}(x) \simeq \pm \frac{ 0.077\,g_{\rm eff}^2\,f\,\tilde{\sigma}_0}{g_*^2 (n+1)} \left(\frac{1}{(x_i^\pm)^{n+1}}  - \frac{1}{x^{n+1}} \right) +\tilde{Y}_{i\pm}.     
\end{equation} 
where we have used $\tilde{Y}_{\rm EQ}\simeq0.278\,g_{\rm eff}/g_*$ with $g_{\rm eff} = 3g_\chi/4$ for fermion and neglected $\tilde{Y}_{\pm}^2$ for the out-of-chemical equilibrium production. As it undergoes weak freeze-out process, all abundance of DM is preserved during and after the freeze-out process,
\begin{eqnarray}
\tilde{Y}_f=Y_+(x\gg x_b) \simeq 2 \tilde{Y}_-(x_b)~,\qquad \tilde{Y}_-(x_b) \simeq  \frac{0.102\, g_{\rm eff}^2 g_*^{-3/2}m_\chi M_{p}\,\tilde{\sigma}_0}{(n+1)\,x_b^{n+1}}. \label{eq:yffh} 
\end{eqnarray}
Substituting our result, Eq.~\ref{eq:yffh}, into the current observational constraint, Eq.~\ref{eq:ocyt}, a  characteristic relation of $m_\chi$, $ \tilde{\sigma}_0 $ and $T_b$ can be obtained. Some examples are shown in Fig.~\ref{fig:fig_high_a-x-Omega-eps-converted-to.pdf}
and in Fig.~\ref{fig:fig_high_a-x-m-eps-converted-to.pdf}, 
where we choose $g_*=90$, $g_\chi=g_{\rm eff}=1$ and $n=2$.

To sum up the discussion on the fermionic DM in a high temperature bounce, we give the criteria for the 
out-of-chemical equilibrium production of 
DM~\cite{Cheung:2014nxi}. 
By rewriting the Boltzmann equation as
\begin{equation}
\frac{x}{\tilde{Y}}\frac{d\tilde{Y}}{dx}= \frac{n_{\chi} \widetilde{\langle\sigma v\rangle}}{H}\left(\frac{\tilde{Y}_{eq}^2}{Y^2}-1\right)~,
\end{equation}  
it is clear that the sufficient condition for  out-of-chemical equilibrium production (i.e. $Y_{\rm EQ}^2/Y^2>1$) is $n_{\chi} \langle\sigma v\rangle < |H|$.
 Therefore if the DM is produced mainly at some 
 temperature $T_*$ with $\Omega_\chi<1$, the condition becomes 
$10^{10} ({\rm GeV})^2\widetilde{\langle\sigma v\rangle} <x_*$ (cf. (1) in \cite{Chung:1998ua}).
 Since in our case the dominant contribution comes 
 from $T_b$ (or $x_b$) (see Eq.~\ref{eq:yffh}), 
 the criteria for an out-of-chemical equilibrium 
 production  is 
\begin{equation} \label{NEQ}
10^{10} ({\rm GeV})^2 \tilde{\sigma}_0 <x_b^{n+1}
\end{equation}   
in this case.

 \section{Thermal Fluctuations of Dark Matter in the  Bounce Universe Scenario}
\label{sec:3}
A crucial ingredient of big bounce genesis~(BBG) is that the production of DM is out-of-chemical equilibrium. Therefore, the informations of early universe evolution can be  preserved in the relic abundance of DM, 
which gives  a tell-tale feature  of a  big bounce. 
Moreover, since the abundance of DM is much less than its thermal equilibrium value, the thermal fluctuations of DM can be generated in this way--in contrast to the case 
that the thermal fluctuations are suppressed in a thermal equilibrium production, such as standard WIMP or WIMP-less Miracle scenarios.
 Empirically, much more information of the early universe evolution must be encoded in the thermal fluctuations
  of DM comparing with that encoded in its relic abundance. Therefore, we are well motivated to study the 
   thermal fluctuations  of DM in BBG. 

Another crucial ingredient of BBG is that the cosmic evolution of universe is bouncing, which consists of a contraction and an expansion. During the accelerating contraction, the effective horizon of universe shrinks so that the wavelength of the long wavelength mode of thermal fluctuation of DM becomes larger than the effective horizon, i.e. being super-horizon. Therefore, only the sub-horizon mode of thermal fluctuation of DM can be investigated by utilizing the conventional thermodynamics approach in which the background of statistical system is assumed to be (quasi-)static~\cite{Cai:2009rd, Biswas:2013lna}. To study the evolution of the super-horizon thermal perturbation modes requires a method beyond the conventional thermodynamics. 

Recently, an integrated scheme is proposed to investigate the evolution of both super-horizon and sub-horizon modes thermal perturbation modes of DM in the contracting and expanding phase of a generic bounce universe~\cite{Li:2015egy}. 

This scheme consists of the following four steps.
\begin{itemize}
\item Step I: {\bf (Inside Horizon)} Computing the energy density of the sub-horizon modes of thermal fluctuations of DM, $\delta\rho_k|_{k\ge |aH|}$, by utilizing the traditional thermodynamics;  

The energy density of sub-horizon thermal fluctuation takes~\cite{Cai:2009rd, Biswas:2013lna}
\begin{equation}
\delta\rho_L^2=\frac{\langle\delta E_\chi\rangle^2}{(aL)^6}=\frac{\mu^2 e^{\beta\mu}}{\pi^{2}\beta^3}(aL)^{-3}~, \quad  aL\le |H|^{-1}~,
\end{equation}  
where $\delta\rho_L$ is short for $\langle\delta\rho_\chi\rangle_{aL}$ with subscript $aL$ denoting the physical length of the given volume, $\mu$ is chemical potential, and $\langle\delta E_\chi\rangle^2$ is thermal fluctuation of DM in the given sub-horizon volume. The $L$-dependence of $\delta\rho_L$ implies the distribution of amplitude for thermal fluctuation modes for each wavelength, which empowers us to go from the real space $L$ to the momentum space $k$ and obtain the power spectrum of the thermal fluctuations for all sub-horizon mode, 
\begin{equation}\label{eq:rksb}
\delta\rho_{k}^2=\frac{6\pi^2\delta \rho_L^2}{k^3}=\frac{6\mu^2 e^{\beta\mu}}{(a\beta)^3}~,\quad k\ge |aH|
\end{equation}
Notice that Eq.~\ref{eq:rksb} is only valid for the sub-horizon modes, and the super-horizon modes is discussed at following. 
\\

\item Step II: {\bf (Beyond Horizon)} Getting the solution of the energy density of the super-horizon modes of thermal perturbations, $\delta\rho_k|_{k\le |aH|}$, by deriving and solving their equation of motion in long wavelength limit, and leaving the initial amplitude of these long wavelength perturbations undetermined; 

Being different from the sub-horizon thermal fluctuations originating from the thermal uncertainties and correlations in the grand ensemble, the super-horizon thermal perturbations describes how the energy density varies with the spatial variance of underlying physical quantities, such as local temperature and chemical potential. So the start point for investigating super-horizon mode of DM thermal fluctuation can be taken as 
\begin{equation}\label{eq:drsp}
\delta\hat{\rho}_\chi({\rm x},t)=\delta n_\chi({\rm x},t)\epsilon_\chi(t)+n_\chi(t)\delta\epsilon_\chi({\rm x},t)~,
\end{equation}  
where the $\hat{}$ on $\delta\rho$ denotes the super-horizon mode, $\epsilon_\chi\equiv \langle E_\chi\rangle/N_\chi$ is the average energy for one DM particle. Without loss of generality, we attribute all such thermal perturbation into the perturbation of temperature, 
\begin{equation}
\tilde{\beta}=\beta+\delta\beta({\rm x},t)~, \qquad \delta \mu({\rm x},t)=0~,
\end{equation}
and obtain
\begin{equation}\label{eq:dtrrb}
\delta\hat{\rho}_\chi({\rm x},t)=-n_\chi\mu\left(\mu-3\beta^{-1}\right)\delta\beta({\rm x},t)~,
\end{equation} 
where ${\rm x}$ denotes spatial coordinates, $\beta\equiv T^{-1}$, $\tilde{A}$ includes the fluctuation and mean value of $A$, $\tilde{A}=A+\delta A({\rm x},t)$, and $\delta A$ is short for $\delta A({\rm x},t)$. 

It is clear that if $\delta\beta({\rm x},t)$ is determined, one can figure out $\delta\hat{\rho}_\chi({\rm x},t)$  with Eq.~\ref{eq:dtrrb} immediately.  By expanding Boltzmann Equation, Eq.~\ref{eq:boe}, up to the first order, and simplifying it with the relation $\partial (\delta\beta)/\partial t=Hy~\partial (\delta \beta)/\partial x$ in radiation dominated background, we can obtain
\begin{equation} \label{eq:dbdx}
\frac{\partial (\delta \beta)}{\partial x}+\frac{\Theta}{xH}\delta \beta=0 ~.
\end{equation}
where $\Theta$ is defined as 
\begin{eqnarray}
 \Theta\equiv \left\{e^{-g(x)}\widetilde{\langle\sigma v\rangle}\frac{m_\chi^3}{\pi^2x^3}\left[1+e^{2g(x)} +6(g(x)-3)^{-1}\right]-[1-\frac{dg(x)}{dx} x(g(x)-3)^{-1}]H\right\}
\end{eqnarray}
for short, $g(y)\equiv \ln\left(n_\chi/n_\chi^{eq}\right)$, $x= m_\chi/T$ for reminder, and the spatial derivative term $\frac{1}{yH}\frac{d {\rm x}^j}{d t}\frac{\partial (\delta \beta)}{\partial {\rm x}^j}$ is discarded in long wavelength limit. To sum up, by solving Eq.~\ref{eq:dbdx} with the abundance of DM in each BBG models, the super-horizon thermal perturbations, $\delta\hat{\rho}_\chi$, is, then, determined by Eq.~\ref{eq:dtrrb}.

At here, we focus on the {\bf Type I} model of BBG, {\it i.e.} bosonic DM in a high temperature bounce, for illustration. From Eq.~\ref{eq:yev}, we have 
\begin{equation}\label{eq:rbi}
n_\chi=\frac{1-e^{-\Lambda(1\mp x)}}{1+e^{-\Lambda(1\mp x)}}n_\chi^{eq}~,\quad \Lambda\equiv 2\pi^2f\langle\sigma v\rangle m_\chi~,
\end{equation}
during the pre-bounce contraction($-$) and the post-bounce expansion($+$). By substituting Eq.~\ref{eq:rbi} into Eq.~\ref{eq:dbdx} and Eq.~\ref{eq:dtrrb} and solving them in high temperature limit, $x\ll 1$,  the evolution of super-horizon modes of thermal perturbation of DM are obtained, 
\begin{equation}\label{eq:rspe}
\delta\hat{\rho}(t)=\left(\frac{\beta(t_i^\mp)}{\beta(t)}\right)^4\delta\hat{\rho}(t_i^\mp)\quad \Longrightarrow \quad \delta\hat{\rho}_k(t)=\left(\frac{\beta(t_i^\mp)}{\beta(t)}\right)^4\delta\hat{\rho}_{k}(t_i^\mp)~.
\end{equation}
where $t_i$ is initial time that super-horizon modes are generated, and we take Fourier transformation at the last step. 
\\
\item Step III: {\bf (Matching on Horizon Crossing)} During the contraction of universe, the effective horizon $|aH|^{-1}$ shrinks, so that the previously sub-horizon modes will becomes super-horizon after horizon crossing. Then the sub-horizon mode and the super-horizon mode can be matched on the moment of horizon crossing, $k=|aH|$, to determine the initial amplitude of the super-horizon thermal perturbations. Afterwards, the evolution of super-horizon thermal perturbation are fully determined during the contacting phase;

The sub-horizon modes with $k$ crosses the effective horizon at different time. And the horizon crossing condition is,
\begin{equation}\label{eq:hcatti}
k=|aH|_{t=t_i^-}~,
\end{equation} 
which leads $\beta(t_i^-)=\frac{\mathcal{C}_0}{4\pi^2f}k^{-1}$ with $\mathcal{C}_0\equiv a(t_i^-)/\beta(t_i^-)=0.752\times10^{-5}\text{eV}$. 

After horizon crossing, each sub-horizon mode becomes super-horizon. So the initial value of each super-horizon mode is determined by the value of sub-horizon mode at horizon crossing,
\begin{equation}\label{eq:rkti}
\delta\hat{\rho}_{k}^2(t_i^-)=\left.\delta\rho_{k}^2\right|_{k= |aH|}=\left.\frac{6\mu^2 e^{\beta\mu}}{(a\beta)^3}\right|_{t=t_i^-}~,
\end{equation}   
where we have used Eq.(\ref{eq:rksb}) with taking $k=|aH|$ at $t=t_i^-$~. 

Substituting these two matching conditions, Eq.(\ref{eq:hcatti}) and Eq.(\ref{eq:rkti}), on the horizon crossing into  Eq.(\ref{eq:rspe}), the evolution of super-horizon mode of thermal perturbation in the contracting phase is obtained,
\begin{eqnarray}\label{eq:rkficp}
 \delta\hat{\rho}_k(t)=\left\{  
\begin{array} {l}
 {\displaystyle \left(\frac{1}{\beta(t)}\right)^4 \frac{\sqrt{6}}{\mathcal{C}_0^\frac{3}{2}}\frac{\Lambda}{2} \ln\left(\frac{2}{\Lambda}\right),~\Lambda\ll 1}   \\ 
 \\ 
  {\displaystyle\left(\frac{1}{\beta(t)}\right)^4 \frac{\sqrt{6}}{\mathcal{C}_0^\frac{3}{2}} 2e^{-\Lambda}, ~\qquad \quad\Lambda\gg 1}    \\
\end{array}     
\right. .
\end{eqnarray}

\item Step IV: {\bf (Matching on Bounce Point)} Eventually, universe are bouncing from the contracting phase to the expanding phase. By assuming the entropy of cosmological background are conserved before and after bounce point, the matching conditions at the bounce point are obtained. By utilizing these matching condition, the evolution of super-horizon thermal perturbation can be also fully determined during the expanding phase.

By assuming the entropy of the bounce are conserved~\cite{Cai:2011ci}, we have an additional pair of matching condition on the bounce,
\begin{equation}\label{eq:bomacob}
\beta(t_f^-)=\beta(t_i^+)~,\qquad \delta\hat{\rho}_k(t_f^-)=\delta\hat{\rho}_k(t_i^+)~,
\end{equation}
where $t_f^-$ is the moment of the contracting phase ending. Again, substituting Eq.(\ref{eq:bomacob}) and  Eq.(\ref{eq:rkficp})  into Eq.(\ref{eq:rspe}),  the evolution of super-horizon mode of thermal perturbation during the expanding phase is fully determined,
\begin{eqnarray}\label{eq:rkfiep}
 \delta\hat{\rho}_k(t)=\left\{  
\begin{array} {l}
 {\displaystyle \left(\frac{1}{\beta(t)}\right)^4 \frac{\sqrt{6}}{\mathcal{C}_0^\frac{3}{2}}\frac{\Lambda}{2} \ln\left(\frac{2}{\Lambda}\right),~\Lambda\ll 1}   \\ 
 \\ 
  {\displaystyle\left(\frac{1}{\beta(t)}\right)^4 \frac{\sqrt{6}}{\mathcal{C}_0^\frac{3}{2}} 2e^{-\Lambda}, ~\qquad \quad \Lambda\gg 1}    \\
\end{array}     
\right. .
\end{eqnarray}

\end{itemize}

With realization of these four steps in detail following~\cite{Li:2015egy}, the energy density spectra of thermal fluctuations of DM in BBG is obtained. It turns out that the amplitude of thermal perturbation of DM are dependent on the particle nature of DM, {\it i.e.} the value of $\Lambda\equiv 2\pi^2f\langle\sigma v\rangle m_\chi$. Moreover the two avenues of DM evolution in the bounce cosmology, {\bf Route I} and {\bf Route II}, can be distinguished in the level of thermal perturbation.  Such results are, hopefully, to be applied to the issues of formation of large scale structure as well as the primordial black hole in near future study~\cite{Navarro:1995iw, Zhao:2002rm,Diemand:2005vz,Frenk:2012ph,Bromm:2009uk, Umeda:2009yc, Colberg:2000zv,DiMatteo:2003zx, Maccio':2012uh}. Moreover, in BBG, such predictions from thermal fluctuation can be used to cross-check with the prediction from direct detection of DM.


\section{Direct Detections of Dark Matter to test the  Big Bounce Genesis}
\label{sec:4}

In order to unravel the nature of DM, it is essential to directly detect it. The possibility of such direct detection, of course, depends on the nature of the DM constituents and their interactions. For the extremely non-relativistic DM candidates such as WIMP and most species of DM candidates in BBG, their average kinetic energy at today are too low to excite the nucleus. So they can be directly detected mainly via the recoiling of a nucleus $(A,Z)$ in elastic scattering. The event rate for such a process is mainly determined with the following three ingredients~\cite{LS96}: 
\begin{enumerate}
\item The elementary DM-nucleon cross section computed in quantum field theory;\\
\item The knowledge of the relevant nuclear matrix elements obtained with as reliable as possible many body nuclear wave functions;\\
\item The knowledge of the density of DM in our vicinity and its velocity distribution.
\end{enumerate}
where the last two ingredients have been discussed extensively in~\cite{Cheung:2014pea}, and we are not, however, going to discuss further these two ingredients in this work.

For the purpose of comparing the predictions of BBG with direct DM search~\cite{XENON10, XENON100.11,XENON10012, CoGeNT11, DAMA1,DAMA11, LUX11, CDMSII04, CRESST, CRESSTII15, PICASSO09,PICASSO11}, we focus on the study of the  DM-nucleon cross section  in this section following~\cite{Cheung:2014pea,Vergados:2016niz}.

Cosmologically, the characteristic relation of the thermally averaged cross section $\langle \sigma v\rangle$ and $m_\chi$ such as Eq.~\ref{eq:macsbbg} have been obtained for the case of the scalar DM particles $\chi$ interact with another scalar $\phi$ via a quartic coupling. However, this relation cannot be utilized directly for comparing with results of the direct detection, because $\langle\sigma v\rangle $ is, essentially, the cross section of the DM pair-annihilation, but not the DM-nucleon cross section, $\sigma_p$, measured in direct detection experiments ~\cite{XENON10, XENON100.11,XENON10012, CoGeNT11, DAMA1,DAMA11, LUX11, CDMSII04, CRESST, CRESSTII15, PICASSO09,PICASSO11}. 

\begin{figure}[htp!]
\centering
\includegraphics[width=0.6\textwidth]{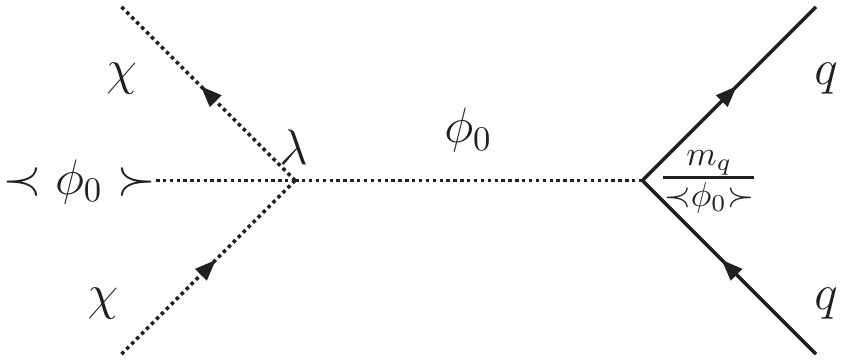}
\caption{The quark - scalar DM scattering mediated by a scalar particle~\cite{Cheung:2014pea}.  }
\label{fig:xxphiphiq}
\end{figure}

The elementary DM-nucleon cross section, $\sigma_p$, can be computed with the Feynman diagram, ~Fig. \ref{fig:xxphiphiq}, in which scalar DM particle interacts with quark mediated by $\phi$~\cite{Cheung:2014pea}.
\\
The resulting DM-nucleon cross section is given by~\cite{Cheung:2014pea}:
 \begin{equation}
  \sigma_p=\frac{1}{4 \pi}\frac{ \lambda^2 m_p^2 (\mu_r^2)}{m_{\phi}^4}\frac{1}{m_{\chi}^2}(\sum_q f_q)^2=\frac{1}{4 \pi}\frac{ \lambda^2 m_p^2}{m_{\phi}^4}\frac{1}{\left (1+m_{\chi}/m_p\right )^2}(\sum_q f_q)^2
  \end{equation}
where $m_p$, $m_\phi$ and $m_\chi$ are, respectively, the mass of proton, $\phi$ and $\chi$. $\mu_r$ is the DM-nucleus reduced mass, and $f_q$ is related to the probability of finding the quark $q$ in the nucleon. 

\begin{figure}[h!t]
\begin{center}
\rotatebox{90}{\hspace{0.0cm} $\sigma_p\rightarrow$pb}
\includegraphics[scale=0.7]{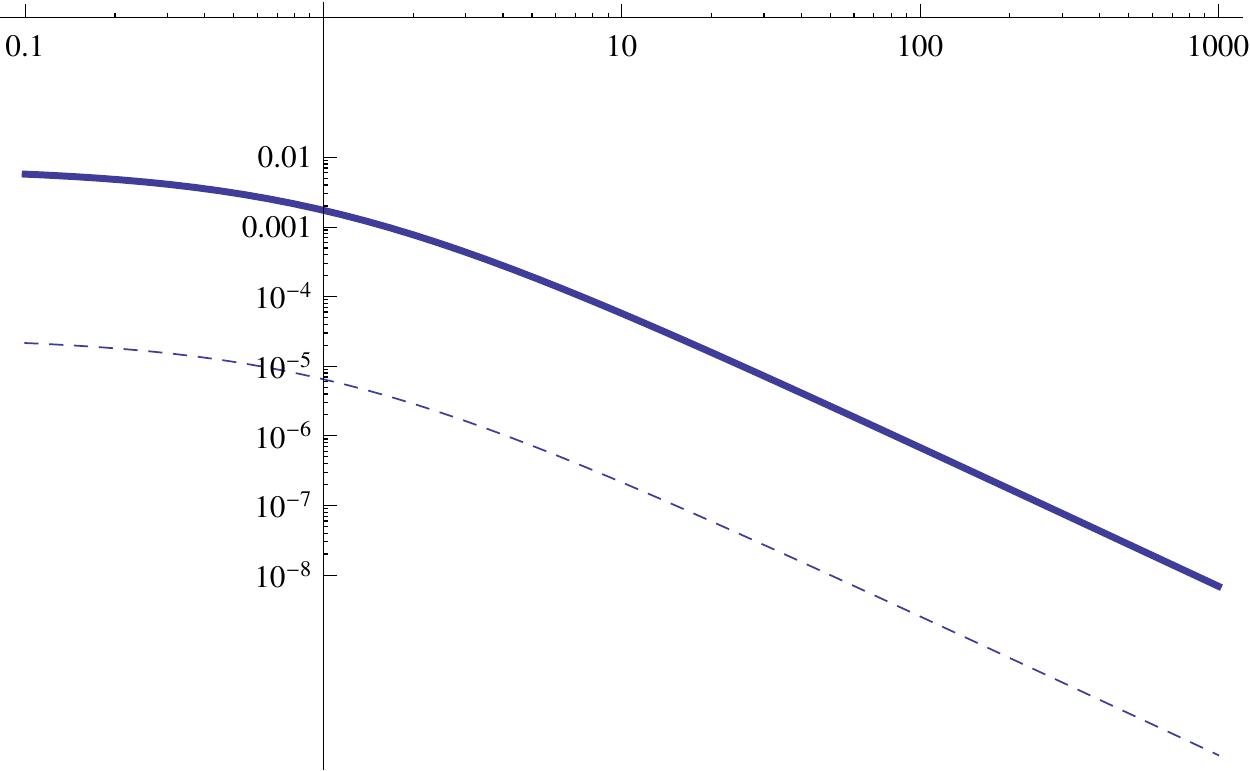}
{\hspace{0.0cm}$m_{\chi} \rightarrow$GeV }
\caption{The nucleon cross section as a function of the DM mass in the case a scalar DM particle: as predicted by our model (thick solid line), consistent with the low DM mass  CRESSTII experiment \cite{CRESSTII15}, and with  the quartic coupling $\lambda$ adjusted to fit the limit of Xe100 experiment, for a DM mass of 50 GeV, i.e. $\sigma_p=10^{-8}$pb (dashed line). }
\label{fig:sigmap}
\end{center}
\end{figure}

Note that the vacuum expectation value  $\prec\phi_0\succ$ in the quartic coupling is canceled by  the Yukawa coupling of scalar $\phi$ with the quarks. Therefore, the elementary DM-nucleon cross section, $\sigma_p$, can be fully determined with a set of parameters, $(m_\chi, m_\phi, \lambda)$.  If the quartic coupling of the scalar DM with the Higgs is the same with the usual quartic coupling of the Higgs particle discovered at LHC, $\lambda=1/2$, $m_{\phi}=126$ GeV, one finds:
 \begin{equation}
  \sigma_p= \sigma_0\left(1+\frac{m_{\chi}}{m_p }\right )^{-2}, \, \sigma_0=6\times 10^{-11}m_p^{-2}\left (\sum_q f_q \right )^2
	\label{Eq:sigma0}
	\end{equation}
	 The value of $\sum_q f_q$ , of course, can  vary a great deal \cite{Chen,Dree00,EFO00},\cite{JDV06}, but its value has now become very much constrained by lattice experiments \cite{GTY09}, yielding  $\sum_q f_q= 0.2$, which we will adopt in this work.  Thus we get $\sigma_0=0.8\times 10^{-3}$pb, which is a bit high leading to $\sigma_p=2.7\times10^{-6}$pb at $m_{\chi}=$50 GeV compared to the limit of $\times10^{-8}$pb extracted the Xe experiments XENON100~\cite{XENON10012,XENON100.11}. We can, of course, treat the quartic Higgs coupling $\lambda$ as a phenomenological parameter and adjust so that it yields  $\sigma_0=2.6\times 10^{-5}$pb yielding the value $\sigma_p=1.0\times10^{-8}$pb  and makes our model consistent with the limit extracted from experiments, e.g. XENON100 ~\cite{XENON10012,XENON100.11} for heavy DM candidates, $m_{\chi}=$50 GeV. The thus obtained nucleon cross sections are exhibited in Fig. \ref{fig:sigmap}. Our model, however, yields  the value of $\sigma_0=0.6\times10^{-4}$pb for $m_{\chi}=$10 GeV, consistent with the the recent low threshold CRESST experiment \cite{CRESSTII15}, which is perhaps more suitable for low mass DM favored by our model. 
	 
We should stress that the DM nucleon cross section dependence exhibited in the exclusion plots is purely kinematic and it does not contain any actual dependence on the cross section of the elementary nucleon cross section as in our model. The extra  mass dependence of the cross section of scalar DM,  exhibiting an enhancement  in  the low DM mass regime, may favor the searches at low energy transfers. It is interesting to compare the behavior of this cross section with that of the relic abundance of the BBG shown in Fig.~\ref{fig:cbplog}.

\begin{figure}[htp!]
\centering
\includegraphics[width=0.7\textwidth]{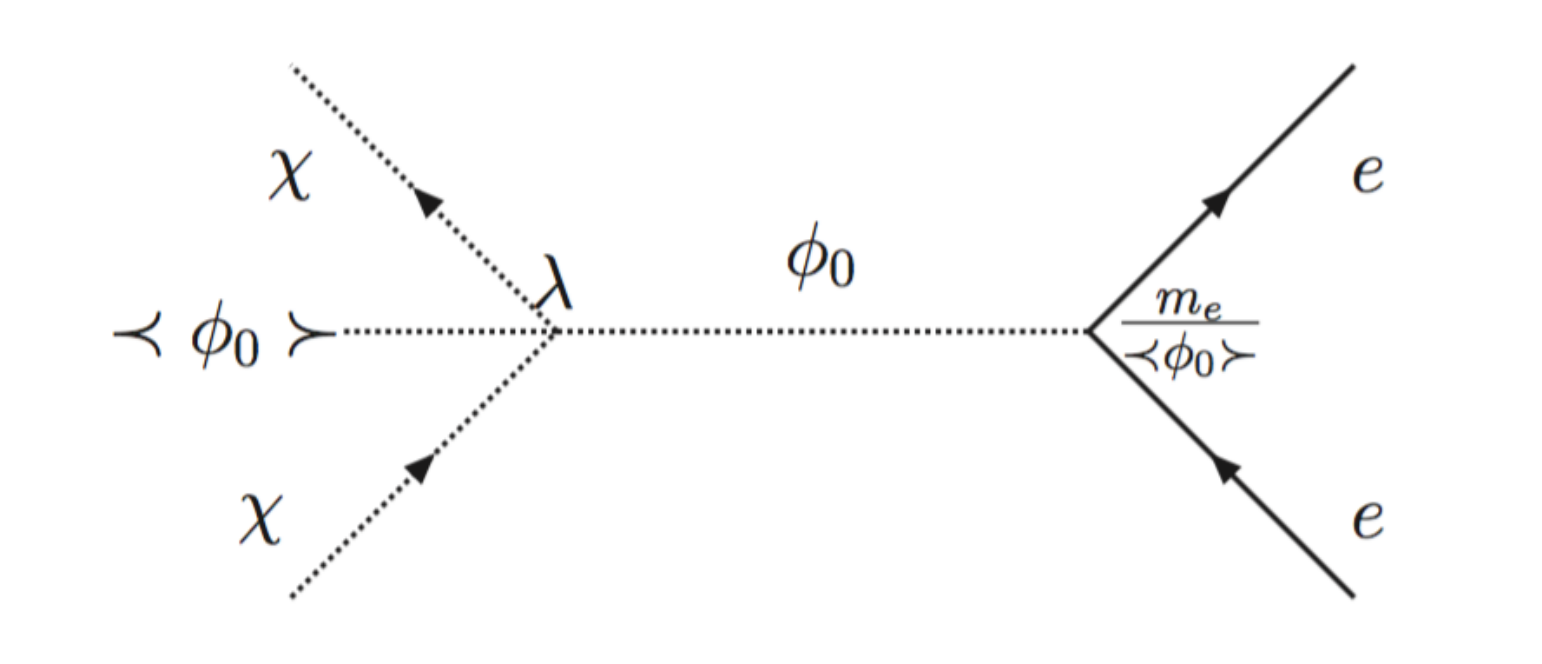}
\caption{The quark - scalar dark matter scattering mediated by a scalar particle~\cite{Cheung:2014pea}.  }
\label{fig:xxphiphie}
\end{figure}

At the end, we also notice another interesting domain of this BBG model in which DM candidate is light. One can find that such DM with mass less than 100 MeV  cannot produce  a detectable recoiling nucleus, but they could produce electrons \cite{MVE05}  with energies in the tens of eV, which  could be detected with current mixed phase detectors ~\cite{XENON14}. If the DM is a scalar particle, however, it can interact in a similar pattern with other fermions, e.g. electrons. The  relevant Feynman diagram is shown in Fig. \ref{fig:xxphiphie}.
\\

For DM with mass in the range of  the electron mass, both the DM particle and the electron are not relativistic. So the expression for  elementary electron cross section is similar to that of hadrons , i.e. it is now given by:
\begin{equation}
  \sigma_e=\frac{1}{4 \pi}\frac{ \lambda^2 m_e^2}{m_{\phi}^4} \left (\frac{m_e m_{\chi}}{m_e+m_{\chi}}\right )^2\frac{1}{m_{\chi}^2}
 \end{equation}
This is a respectable size cross section dependent on the ratio $m_{\chi}/m_e$. In this case one must consider electron recoils, but the highest possible electron energy is about 1.5 eV and the DM mass must greater than 0.3 electron masses. So the detection of  DM with mass around the electron mass requires another type of detector~\cite{Cheung:2014pea, Li:forthcoming, Vergados:2016niz}.


\section{Summary}
In this review  we discuss the big bounce genesis~(BBG) as an unified framework for the  interplay between dark matter~(DM) and early evolution of universe, particularly, in which the famous WIMP and WIMP-less Miracles are also included. 
The novelty of BBG is that it provides a new possibility of 
using DM mass and its predicted interaction cross section, 
as a telling  signal of the existence of a big bounce at the early stage in the evolution of our currently observed universe. Each type of BBG models, {\it bosonic/fermionic DM in a high/low temperature bounce}, and its predictions have been discussed in details.
 These predictions  can be checked  against  data from the 
 present and future DM searches. 

Another salient feature of BBG is that {out-of-chemical equilibrium production} is allowed. In this case, the abundance of DM is much less than its thermal equilibrium value, the thermal fluctuation of DM, then, can be generated; in sharp contrast to the case in which  thermal fluctuations are
 suppressed in the thermal equilibrium production. In this review, we also present a detailed and model-independent 
 analysis of the whole evolution of DM thermal fluctuations 
  in a generic bounce. 
 It may have  important implications on 
  the formation of large scale structure, 
  clusters, galaxies and primordial blackhole, 
  and potentially can be compared with astrophysical observations. 

At the end of this review, we also present detailed analysis
 for the predicted event rates for different DM 
 detection experiments: 
  events/kg/year for  nuclear recoil experiments for 
  heavy scalar DM in the mass range of ~100GeV, and 
  event rate for ~MeV light scalar DM detections by electronic scattering.


\paragraph{\bf Acknowledgments.}
Y.-K.E.C would like to thank Jin U Kang and  Konstantin Savvidy for many useful discussions. 
Y.-K.E.C was supported in part by NSFC grant under contract~11405084. Y.-K.E.C  also acknowledges  985-Grant 
from the Chinese Ministry of Education, and the Priority 
Academic Program Development for Jiangsu Higher Education Institutions (PAPD).\\
C.L. has been supported in parts by the NSFC grants under contract~11603018 and contract~11433004, 
the Young Investigator in Fundamental Research  Grants of the  Yunnan Provincial Ministry of Science and Technology under contract number~2016FD006, 
the Leading Talents of Yunnan Province under contract number~2015HA022 and the Top Talents of Yunnan Province. \\
JDV is indebted to the  ARC Centre of Excellence in Particle Physics at the Terascale and Centre for the Subatomic Structure of Matter (CSSM) at University of Adelaide for their kind invitation and support,  and to Professor Tony Thomas,
 Director of ARC, for for useful discussions and 
 his hospitality.

\clearpage
\addcontentsline{toc}{section}{References}

\bibliographystyle{JHEP}

\bibliography{Tex}

\providecommand{\href}[2]{#2}\begingroup\raggedright\begin{thebibliography}{100}

\bibitem{Laozi}
Laozi, {\em {14th Chapter of Tao Te Ching}}.
\newblock 6th century BC.

\bibitem{Guth:1980zm}
A.~H. Guth, {\it {The Inflationary Universe: A Possible Solution to the Horizon
  and Flatness Problems}},  {\em Phys.Rev.} {\bf D23} (1981) 347--356.

\bibitem{Komatsu:2010fb}
{\bf WMAP Collaboration} Collaboration, E.~Komatsu et~al., {\it {Seven-Year
  Wilkinson Microwave Anisotropy Probe (WMAP) Observations: Cosmological
  Interpretation}},  {\em Astrophys.J.Suppl.} {\bf 192} (2011) 18,
  [\href{http://xxx.lanl.gov/abs/1001.4538}{{\tt arXiv:1001.4538}}].

\bibitem{Ade:2013kta}
{\bf Planck} Collaboration, P.~Ade et~al., {\it {Planck 2013 results. XV. CMB
  power spectra and likelihood}},  {\em Astron.Astrophys.} {\bf 571} (2014)
  A15, [\href{http://xxx.lanl.gov/abs/1303.5075}{{\tt arXiv:1303.5075}}].

\bibitem{Mukhanov:1990me}
V.~F. Mukhanov, H.~Feldman, and R.~H. Brandenberger, {\it {Theory of
  cosmological perturbations. Part 1. Classical perturbations. Part 2. Quantum
  theory of perturbations. Part 3. Extensions}},  {\em Phys.Rept.} {\bf 215}
  (1992) 203--333.

\bibitem{Borde:1993xh}
A.~Borde and A.~Vilenkin, {\it {Eternal inflation and the initial
  singularity}},  {\em Phys.Rev.Lett.} {\bf 72} (1994) 3305--3309,
  [\href{http://xxx.lanl.gov/abs/gr-qc/9312022}{{\tt gr-qc/9312022}}].

\bibitem{NoBer08}
M.~Novello and S.~P. Bergliaffa {\em Phys. Rep.} {\bf 463} (2008) 127. arXiv:
  0802.1634 [astro-ph].

\bibitem{Branden12}
R. H. Brandenberger, (2012), arXiv:1206.4196 [astro-ph.CO].

\bibitem{Battefeld:2014uga}
D.~Battefeld and P.~Peter, {\it {A Critical Review of Classical Bouncing
  Cosmologies}},  {\em Phys. Rept.} {\bf 571} (2015) 1--66,
  [\href{http://xxx.lanl.gov/abs/1406.2790}{{\tt arXiv:1406.2790}}].

\bibitem{Brandenberger:2016vhg}
R.~Brandenberger and P.~Peter, {\it {Bouncing Cosmologies: Progress and
  Problems}},  \href{http://xxx.lanl.gov/abs/1603.0583}{{\tt arXiv:1603.0583}}.

\bibitem{Khoury:2001wf}
J.~Khoury, B.~A. Ovrut, P.~J. Steinhardt, and N.~Turok, {\it {The Ekpyrotic
  universe: Colliding branes and the origin of the hot big bang}},  {\em
  Phys.Rev.} {\bf D64} (2001) 123522,
  [\href{http://xxx.lanl.gov/abs/hep-th/0103239}{{\tt hep-th/0103239}}].

\bibitem{Steinhardt:2001st}
P.~J. Steinhardt and N.~Turok, {\it {Cosmic evolution in a cyclic universe}},
  {\em Phys. Rev.} {\bf D65} (2002) 126003,
  [\href{http://xxx.lanl.gov/abs/hep-th/0111098}{{\tt hep-th/0111098}}].

\bibitem{Steinhardt:2002ih}
P.~J. Steinhardt and N.~Turok, {\it {A cyclic model of the universe}},  {\em
  Science} {\bf 296} (2002) 1436--1439.

\bibitem{Gasperini:2002bn}
M.~Gasperini and G.~Veneziano, {\it {The Pre - big bang scenario in string
  cosmology}},  {\em Phys.Rept.} {\bf 373} (2003) 1--212,
  [\href{http://xxx.lanl.gov/abs/hep-th/0207130}{{\tt hep-th/0207130}}].

\bibitem{Creminelli:2006xe}
P.~Creminelli, M.~A. Luty, A.~Nicolis, and L.~Senatore, {\it {Starting the
  Universe: Stable Violation of the Null Energy Condition and Non-standard
  Cosmologies}},  {\em JHEP} {\bf 0612} (2006) 080,
  [\href{http://xxx.lanl.gov/abs/hep-th/0606090}{{\tt hep-th/0606090}}].

\bibitem{Cai:2007qw}
Y.-F. Cai, T.~Qiu, Y.-S. Piao, M.~Li, and X.~Zhang, {\it {Bouncing universe
  with quintom matter}},  {\em JHEP} {\bf 0710} (2007) 071,
  [\href{http://xxx.lanl.gov/abs/0704.1090}{{\tt arXiv:0704.1090}}].

\bibitem{Cai:2008qw}
Y.-F. Cai, T.-t. Qiu, R.~Brandenberger, and X.-m. Zhang, {\it {A Nonsingular
  Cosmology with a Scale-Invariant Spectrum of Cosmological Perturbations from
  Lee-Wick Theory}},  {\em Phys.Rev.} {\bf D80} (2009) 023511,
  [\href{http://xxx.lanl.gov/abs/0810.4677}{{\tt arXiv:0810.4677}}].

\bibitem{Wands:2008tv}
D.~Wands, {\it {Cosmological perturbations through the big bang}},  {\em
  Adv.Sci.Lett.} {\bf 2} (2009) 194--204,
  [\href{http://xxx.lanl.gov/abs/0809.4556}{{\tt arXiv:0809.4556}}].

\bibitem{Bhattacharya:2013ut}
K.~Bhattacharya, Y.-F. Cai, and S.~Das, {\it {Lee-Wick radiation induced
  bouncing universe models}},  {\em Phys.Rev.} {\bf D87} (2013), no.~8 083511,
  [\href{http://xxx.lanl.gov/abs/1301.0661}{{\tt arXiv:1301.0661}}].

\bibitem{Odintsov:2014gea}
S.~D. Odintsov and V.~K. Oikonomou, {\it {Matter Bounce Loop Quantum Cosmology
  from $F(R)$ Gravity}},  {\em Phys. Rev.} {\bf D90} (2014), no.~12 124083,
  [\href{http://xxx.lanl.gov/abs/1410.8183}{{\tt arXiv:1410.8183}}].

\bibitem{Li:2014msi}
H.~Li, M.~Li, T.~Qiu, J.~Xia, Y.~Piao, et~al., {\it {What can we learn from the
  tension between PLANCK and BICEP2 data?}},  {\em Sci.China Phys.Mech.Astron.}
  {\bf 57} (2014) 1431--1441.

\bibitem{Quintin:2014oea}
J.~Quintin, Y.-F. Cai, and R.~H. Brandenberger, {\it {Matter Creation in a
  Nonsingular Bouncing Cosmology}},
  \href{http://xxx.lanl.gov/abs/1406.6049}{{\tt arXiv:1406.6049}}.

\bibitem{Wan:2014fra}
Y.~Wan, S.~Li, M.~Li, T.~Qiu, Y.~Cai, et~al., {\it {Single field inflation with
  modulated potential in light of the Planck and BICEP2}},
  \href{http://xxx.lanl.gov/abs/1405.2784}{{\tt arXiv:1405.2784}}.

\bibitem{Cai:2014bea}
Y.-F. Cai, {\it {Exploring Bouncing Cosmologies with Cosmological Surveys}},
  {\em Sci.China Phys.Mech.Astron.} {\bf 57} (2014) 1414--1430,
  [\href{http://xxx.lanl.gov/abs/1405.1369}{{\tt arXiv:1405.1369}}].

\bibitem{Liu:2014tda}
Z.-G. Liu, H.~Li, and Y.-S. Piao, {\it {Pre-inflationary genesis with CMB
  B-mode polarization}},  \href{http://xxx.lanl.gov/abs/1405.1188}{{\tt
  arXiv:1405.1188}}.

\bibitem{Li:2014qwa}
M.~Li, {\it {Generating scale-invariant tensor perturbations in the
  non-inflationary universe}},  \href{http://xxx.lanl.gov/abs/1405.0211}{{\tt
  arXiv:1405.0211}}.

\bibitem{Cai:2014hja}
Y.-F. Cai and Y.~Wang, {\it {Testing quantum gravity effects with latest CMB
  observations}},  {\em Phys.Lett.} {\bf B735} (2014) 108--111,
  [\href{http://xxx.lanl.gov/abs/1404.6672}{{\tt arXiv:1404.6672}}].

\bibitem{Cai:2014xxa}
Y.-F. Cai, J.~Quintin, E.~N. Saridakis, and E.~Wilson-Ewing, {\it {Nonsingular
  bouncing cosmologies in light of BICEP2}},
  \href{http://xxx.lanl.gov/abs/1404.4364}{{\tt arXiv:1404.4364}}.

\bibitem{Hu:2014aua}
B.~Hu, J.-W. Hu, Z.-K. Guo, and R.-G. Cai, {\it {Reconstruction of the
  primordial power spectra with Planck and BICEP2}},
  \href{http://xxx.lanl.gov/abs/1404.3690}{{\tt arXiv:1404.3690}}.

\bibitem{Li:2014cka}
H.~Li, J.-Q. Xia, and X.~Zhang, {\it {Global fitting analysis on cosmological
  models after BICEP2}},  \href{http://xxx.lanl.gov/abs/1404.0238}{{\tt
  arXiv:1404.0238}}.

\bibitem{Xia:2014tda}
J.-Q. Xia, Y.-F. Cai, H.~Li, and X.~Zhang, {\it {Evidence for bouncing
  evolution before inflation after BICEP2}},  {\em Phys.Rev.Lett.} {\bf 112}
  (2014) 251301, [\href{http://xxx.lanl.gov/abs/1403.7623}{{\tt
  arXiv:1403.7623}}].

\bibitem{Nojiri:2015sfd}
S.~Nojiri, S.~D. Odintsov, and V.~K. Oikonomou, {\it {Unimodular $F(R)$
  Gravity}},  {\em JCAP} {\bf 1605} (2016), no.~05 046,
  [\href{http://xxx.lanl.gov/abs/1512.0722}{{\tt arXiv:1512.0722}}].

\bibitem{Nojiri:2016ygo}
S.~Nojiri, S.~D. Odintsov, and V.~K. Oikonomou, {\it {Bounce universe history
  from unimodular $F(R)$ gravity}},  {\em Phys. Rev.} {\bf D93} (2016), no.~8
  084050, [\href{http://xxx.lanl.gov/abs/1601.0411}{{\tt arXiv:1601.0411}}].

\bibitem{Cheung:2016oab}
Y.-K.~E. Cheung, X.~Song, S.~Li, Y.~Li, and Y.~Zhu, {\it {A smoothly bouncing
  universe from String Theory}},  \href{http://xxx.lanl.gov/abs/1601.0380}{{\tt
  arXiv:1601.0380}}.

\bibitem{Escofet:2015gpa}
A.~Escofet and E.~Elizalde, {\it {Gauss?Bonnet modified gravity models with
  bouncing behavior}},  {\em Mod. Phys. Lett.} {\bf A31} (2016), no.~17
  1650108, [\href{http://xxx.lanl.gov/abs/1510.0584}{{\tt arXiv:1510.0584}}].

\bibitem{Haro:2015oqa}
J.~Haro, A.~N. Makarenko, A.~N. Myagky, S.~D. Odintsov, and V.~K. Oikonomou,
  {\it {Bouncing loop quantum cosmology in Gauss-Bonnet gravity}},  {\em Phys.
  Rev.} {\bf D92} (2015), no.~12 124026,
  [\href{http://xxx.lanl.gov/abs/1506.0827}{{\tt arXiv:1506.0827}}].

\bibitem{Odintsov:2016tar}
S.~D. Odintsov and V.~K. Oikonomou, {\it {Deformed Matter Bounce with Dark
  Energy Epoch}},  \href{http://xxx.lanl.gov/abs/1606.0368}{{\tt
  arXiv:1606.0368}}.

\bibitem{Choudhury:2015baa}
S.~Choudhury and S.~Banerjee, {\it {Hysteresis in the Sky}},  {\em Astropart.
  Phys.} {\bf 80} (2016) 34--89, [\href{http://xxx.lanl.gov/abs/1506.0226}{{\tt
  arXiv:1506.0226}}].

\bibitem{Odintsov:2015ynk}
S.~D. Odintsov and V.~K. Oikonomou, {\it {Big-Bounce with Finite-time
  Singularity: The $F(R)$ Gravity Description}},
  \href{http://xxx.lanl.gov/abs/1512.0478}{{\tt arXiv:1512.0478}}.

\bibitem{Oikonomou:2015qha}
V.~K. Oikonomou, {\it {Singular Bouncing Cosmology from Gauss-Bonnet Modified
  Gravity}},  {\em Phys. Rev.} {\bf D92} (2015), no.~12 124027,
  [\href{http://xxx.lanl.gov/abs/1509.0582}{{\tt arXiv:1509.0582}}].

\bibitem{Cai:2014jla}
Y.-F. Cai and E.~Wilson-Ewing, {\it {A $\Lambda$CDM bounce scenario}},  {\em
  JCAP} {\bf 1503} (2015), no.~03 006,
  [\href{http://xxx.lanl.gov/abs/1412.2914}{{\tt arXiv:1412.2914}}].

\bibitem{Cai:2015vzv}
Y.-F. Cai, F.~Duplessis, D.~A. Easson, and D.-G. Wang, {\it {Searching for a
  matter bounce cosmology with low redshift observations}},  {\em Phys. Rev.}
  {\bf D93} (2016), no.~4 043546,
  [\href{http://xxx.lanl.gov/abs/1512.0897}{{\tt arXiv:1512.0897}}].

\bibitem{Ferreira:2016gfg}
E.~G.~M. Ferreira and R.~Brandenberger, {\it {Holographic Curvature
  Perturbations in a Cosmology with a Space-Like Singularity}},  {\em JCAP}
  {\bf 1607} (2016), no.~07 030, [\href{http://xxx.lanl.gov/abs/1602.0815}{{\tt
  arXiv:1602.0815}}].

\bibitem{Quintin:2015rta}
J.~Quintin, Z.~Sherkatghanad, Y.-F. Cai, and R.~H. Brandenberger, {\it
  {Evolution of cosmological perturbations and the production of
  non-Gaussianities through a nonsingular bounce: Indications for a no-go
  theorem in single field matter bounce cosmologies}},  {\em Phys. Rev.} {\bf
  D92} (2015), no.~6 063532, [\href{http://xxx.lanl.gov/abs/1508.0414}{{\tt
  arXiv:1508.0414}}].

\bibitem{Brandenberger:2016egn}
R.~H. Brandenberger, Y.-F. Cai, S.~R. Das, E.~G.~M. Ferreira, I.~A. Morrison,
  and Y.~Wang, {\it {Fluctuations in a Cosmology with a Space-Like Singularity
  and their Gauge Theory Dual Description}},
  \href{http://xxx.lanl.gov/abs/1601.0023}{{\tt arXiv:1601.0023}}.

\bibitem{Hipolito-Ricaldi:2016kqq}
W.~S. Hipolito-Ricaldi, R.~Brandenberger, E.~G.~M. Ferreira, and L.~L. Graef,
  {\it {Particle Production in Ekpyrotic Scenarios}},
  \href{http://xxx.lanl.gov/abs/1605.0467}{{\tt arXiv:1605.0467}}.

\bibitem{Wan:2015hya}
Y.~Wan, T.~Qiu, F.~P. Huang, Y.-F. Cai, H.~Li, and X.~Zhang, {\it {Bounce
  Inflation Cosmology with Standard Model Higgs Boson}},  {\em JCAP} {\bf 1512}
  (2015), no.~12 019, [\href{http://xxx.lanl.gov/abs/1509.0877}{{\tt
  arXiv:1509.0877}}].

\bibitem{Cai:2014zga}
Y.-F. Cai and E.~Wilson-Ewing, {\it {Non-singular bounce scenarios in loop
  quantum cosmology and the effective field description}},  {\em JCAP} {\bf
  1403} (2014) 026, [\href{http://xxx.lanl.gov/abs/1402.3009}{{\tt
  arXiv:1402.3009}}].

\bibitem{Wands:1998yp}
D.~Wands, {\it {Duality invariance of cosmological perturbation spectra}},
  {\em Phys.Rev.} {\bf D60} (1999) 023507,
  [\href{http://xxx.lanl.gov/abs/gr-qc/9809062}{{\tt gr-qc/9809062}}].

\bibitem{Finelli:2001sr}
F.~Finelli and R.~Brandenberger, {\it {On the generation of a scale invariant
  spectrum of adiabatic fluctuations in cosmological models with a contracting
  phase}},  {\em Phys.Rev.} {\bf D65} (2002) 103522,
  [\href{http://xxx.lanl.gov/abs/hep-th/0112249}{{\tt hep-th/0112249}}].

\bibitem{Li:2013bha}
C.~Li and Y.-K.~E. Cheung, {\it {The scale invariant power spectrum of the
  primordial curvature perturbations from the coupled scalar tachyon bounce
  cosmos}},  {\em JCAP} {\bf 1407} (2014) 008,
  [\href{http://xxx.lanl.gov/abs/1401.0094}{{\tt arXiv:1401.0094}}].

\bibitem{Li:2011nj}
C.~Li, L.~Wang, and Y.-K.~E. Cheung, {\it {Bound to bounce: A coupled
  scalar?tachyon model for a smooth bouncing/cyclic universe}},  {\em Phys.
  Dark Univ.} {\bf 3} (2014) 18--33,
  [\href{http://xxx.lanl.gov/abs/1101.0202}{{\tt arXiv:1101.0202}}].

\bibitem{Li:2014cba}
C.~Li, {\it {Thermally producing and weakly freezing out dark matter in a
  bouncing universe}},  {\em Phys. Rev.} {\bf D92} (2015), no.~6 063513,
  [\href{http://xxx.lanl.gov/abs/1404.4012}{{\tt arXiv:1404.4012}}].

\bibitem{Boyle:2004gv}
L.~A. Boyle, P.~J. Steinhardt, and N.~Turok, {\it {A New duality relating
  density perturbations in expanding and contracting Friedmann cosmologies}},
  {\em Phys.Rev.} {\bf D70} (2004) 023504,
  [\href{http://xxx.lanl.gov/abs/hep-th/0403026}{{\tt hep-th/0403026}}].

\bibitem{Li:2012vi}
C.~Li and Y.-K.~E. Cheung, {\it {Dualities between Scale Invariant and
  Magnitude Invariant Perturbation Spectra in Inflationary/Bouncing Cosmos}},
  \href{http://xxx.lanl.gov/abs/1211.1610}{{\tt arXiv:1211.1610}}.

\bibitem{Li:2014era}
C.~Li, R.~H. Brandenberger, and Y.-K.~E. Cheung, {\it {Big Bounce Genesis}},
  {\em Phys. Rev.} {\bf D90} (2014), no.~12 123535,
  [\href{http://xxx.lanl.gov/abs/1403.5625}{{\tt arXiv:1403.5625}}].

\bibitem{Scherrer:1985zt}
R.~J. Scherrer and M.~S. Turner, {\it {On the Relic, Cosmic Abundance of Stable
  Weakly Interacting Massive Particles}},  {\em Phys.Rev.} {\bf D33} (1986)
  1585.

\bibitem{Feng:2008ya}
J.~L. Feng and J.~Kumar, {\it {The WIMPless Miracle: Dark-Matter Particles
  without Weak-Scale Masses or Weak Interactions}},  {\em Phys.Rev.Lett.} {\bf
  101} (2008) 231301, [\href{http://xxx.lanl.gov/abs/0803.4196}{{\tt
  arXiv:0803.4196}}].

\bibitem{Kolb:1990vq}
E.~W. Kolb and M.~S. Turner, {\it {The Early Universe}},  {\em Front.Phys.}
  {\bf 69} (1990) 1--547.

\bibitem{Gondolo:1990dk}
P.~Gondolo and G.~Gelmini, {\it {Cosmic abundances of stable particles:
  Improved analysis}},  {\em Nucl.Phys.} {\bf B360} (1991) 145--179.

\bibitem{Cheung:2014pea}
Y.-K.~E. Cheung and J.~D. Vergados, {\it {Direct dark matter searches - Test of
  the Big Bounce Cosmology}},  {\em JCAP} {\bf 1502} (2015), no.~02 014,
  [\href{http://xxx.lanl.gov/abs/1410.5710}{{\tt arXiv:1410.5710}}].

\bibitem{Cheung:2014nxi}
Y.-K.~E. Cheung, J.~U. Kang, and C.~Li, {\it {Dark matter in a bouncing
  universe}},  {\em JCAP} {\bf 1411} (2014), no.~11 001,
  [\href{http://xxx.lanl.gov/abs/1408.4387}{{\tt arXiv:1408.4387}}].

\bibitem{Cai:2011ci}
Y.-F. Cai, R.~Brandenberger, and X.~Zhang, {\it {Preheating a bouncing
  universe}},  {\em Phys.Lett.} {\bf B703} (2011) 25--33,
  [\href{http://xxx.lanl.gov/abs/1105.4286}{{\tt arXiv:1105.4286}}].

\bibitem{Vergados:2016niz}
J.~D. Vergados, C.~C. Moustakidis, Y.-K.~E. Cheung, H.~Ejri, Y.~Kim, and
  Y.~Lie, {\it {Light WIMP searches involving electron scattering}},
  \href{http://xxx.lanl.gov/abs/1605.0541}{{\tt arXiv:1605.0541}}.

\bibitem{Li:2015egy}
C.~Li, {\it {Thermal Fluctuations of Dark Matter in Bouncing Cosmology}},  {\em
  JCAP} {\bf 1609} (2016), no.~09 038,
  [\href{http://xxx.lanl.gov/abs/1512.0679}{{\tt arXiv:1512.0679}}].

\bibitem{Peskin:1995ev}
M.~E. Peskin and D.~V. Schroeder, {\em {An Introduction to quantum field
  theory}}.
\newblock 1995.

\bibitem{Capozziello:2007ec}
S.~Capozziello and M.~Francaviglia, {\it {Extended Theories of Gravity and
  their Cosmological and Astrophysical Applications}},  {\em Gen. Rel. Grav.}
  {\bf 40} (2008) 357--420, [\href{http://xxx.lanl.gov/abs/0706.1146}{{\tt
  arXiv:0706.1146}}].

\bibitem{Capozziello:2012ie}
S.~Capozziello and M.~De~Laurentis, {\it {The dark matter problem from f(R)
  gravity viewpoint}},  {\em Annalen Phys.} {\bf 524} (2012) 545--578.

\bibitem{Chung:1998ua}
D.~J. Chung, E.~W. Kolb, and A.~Riotto, {\it {Nonthermal supermassive dark
  matter}},  {\em Phys.Rev.Lett.} {\bf 81} (1998) 4048--4051,
  [\href{http://xxx.lanl.gov/abs/hep-ph/9805473}{{\tt hep-ph/9805473}}].

\bibitem{Cai:2009rd}
Y.-F. Cai, W.~Xue, R.~Brandenberger, and X.-m. Zhang, {\it {Thermal
  Fluctuations and Bouncing Cosmologies}},  {\em JCAP} {\bf 0906} (2009) 037,
  [\href{http://xxx.lanl.gov/abs/0903.4938}{{\tt arXiv:0903.4938}}].

\bibitem{Biswas:2013lna}
T.~Biswas, R.~Brandenberger, T.~Koivisto, and A.~Mazumdar, {\it {Cosmological
  perturbations from statistical thermal fluctuations}},  {\em Phys. Rev.} {\bf
  D88} (2013), no.~2 023517, [\href{http://xxx.lanl.gov/abs/1302.6463}{{\tt
  arXiv:1302.6463}}].

\bibitem{Navarro:1995iw}
J.~F. Navarro, C.~S. Frenk, and S.~D.~M. White, {\it {The Structure of cold
  dark matter halos}},  {\em Astrophys. J.} {\bf 462} (1996) 563--575,
  [\href{http://xxx.lanl.gov/abs/astro-ph/9508025}{{\tt astro-ph/9508025}}].

\bibitem{Zhao:2002rm}
D.~Zhao, H.~Mo, Y.~Jing, and G.~Boerner, {\it {The growth and structure of dark
  matter haloes}},  {\em Mon. Not. Roy. Astron. Soc.} {\bf 339} (2003) 12--24,
  [\href{http://xxx.lanl.gov/abs/astro-ph/0204108}{{\tt astro-ph/0204108}}].

\bibitem{Diemand:2005vz}
J.~Diemand, B.~Moore, and J.~Stadel, {\it {Earth-mass dark-matter haloes as the
  first structures in the early Universe}},  {\em Nature} {\bf 433} (2005)
  389--391, [\href{http://xxx.lanl.gov/abs/astro-ph/0501589}{{\tt
  astro-ph/0501589}}].

\bibitem{Frenk:2012ph}
C.~S. Frenk and S.~D.~M. White, {\it {Dark matter and cosmic structure}},  {\em
  Annalen Phys.} {\bf 524} (2012) 507--534,
  [\href{http://xxx.lanl.gov/abs/1210.0544}{{\tt arXiv:1210.0544}}].

\bibitem{Bromm:2009uk}
V.~Bromm, N.~Yoshida, L.~Hernquist, and C.~F. McKee, {\it {The formation of the
  first stars and galaxies}},  {\em Nature} {\bf 459} (2009) 49--54,
  [\href{http://xxx.lanl.gov/abs/0905.0929}{{\tt arXiv:0905.0929}}].

\bibitem{Umeda:2009yc}
H.~Umeda, N.~Yoshida, K.~Nomoto, S.~Tsuruta, M.~Sasaki, and T.~Ohkubo, {\it
  {Early Black Hole Formation by Accretion of Gas and Dark Matter}},  {\em
  JCAP} {\bf 0908} (2009) 024, [\href{http://xxx.lanl.gov/abs/0908.0573}{{\tt
  arXiv:0908.0573}}].

\bibitem{Colberg:2000zv}
{\bf VIRGO} Collaboration, J.~M. Colberg et~al., {\it {Clustering of galaxy
  clusters in CDM universes}},  {\em Mon. Not. Roy. Astron. Soc.} {\bf 319}
  (2000) 209, [\href{http://xxx.lanl.gov/abs/astro-ph/0005259}{{\tt
  astro-ph/0005259}}].

\bibitem{DiMatteo:2003zx}
T.~Di~Matteo, R.~A.~C. Croft, V.~Springel, and L.~Hernquist, {\it {Black hole
  growth and activity in a lambda CDM universe}},  {\em Astrophys. J.} {\bf
  593} (2003) 56--68, [\href{http://xxx.lanl.gov/abs/astro-ph/0301586}{{\tt
  astro-ph/0301586}}].

\bibitem{Maccio':2012uh}
A.~V. Maccio, O.~Ruchayskiy, A.~Boyarsky, and J.~C. Munoz-Cuartas, {\it {The
  inner structure of haloes in Cold+Warm dark matter models}},  {\em Mon. Not.
  Roy. Astron. Soc.} {\bf 428} (2013) 882--890,
  [\href{http://xxx.lanl.gov/abs/1202.2858}{{\tt arXiv:1202.2858}}].

\bibitem{LS96}
J.~D. Lewin and P.~F. Smith {\em Astropart. Phys.} {\bf 6} (1996) 87.

\bibitem{XENON10}
J. Angle {\it et al}, arXiv:1104.3088 [hep-ph].

\bibitem{XENON100.11}
E.~Aprile et~al. {\em Phys. Rev. Lett.} {\bf 107} (2011) 131302.
  arXiv:1104.2549v3 [astro-ph.CO].

\bibitem{XENON10012}
E.~Aprile et~al. {\em Phys. Rev. Lett.} {\bf 109} (2012) 181301. [XENON100
  Collaboration]; arXiv: 1207.5988 (astro-ph.Co).

\bibitem{CoGeNT11}
C.~Aalseth et~al. {\em Phys. Rev. Lett.} {\bf 106} (2011) 131301. CoGeNT
  collaboration arXiv:10002.4703 [astro-ph.CO].

\bibitem{DAMA1}
R.~Bernabei and Others {\em Eur. Phys. J. C} {\bf 56} (2008) 333. [DAMA
  Collaboration]; [arXiv:0804.2741 [astro-ph]].

\bibitem{DAMA11}
P. Belli {\it et al}, arXiv:1106.4667 [astro-ph.GA].

\bibitem{LUX11}
D.C. Malling {it et al}, arXiv:1110.0103((astro-ph.IM)).

\bibitem{CDMSII04}
D.~S.~A. others {\em Phys.Rev.Lett.} {\bf 93} (2004) 211301. the CDMS
  Collaboration.

\bibitem{CRESST}
The CRESST Experiment: Recent Results and Prospects, P.Di Stefano, {\it et al},
  arXiv:hep-ex/0011064; The CRESST Collaboration, talk presented at IBS -
  MultiDark Joint Focus Program, Daejeon, s. Korea, 10 – 21 October 2014.

\bibitem{CRESSTII15}
The CRESST Collaboration, G. Angloher {\it et al}, arXiv:1509.01515
  (astro-ph.CO,astro-ph.IM,physics.ins-det).

\bibitem{PICASSO09}
S.~Archambault et~al. {\em Phys. Lett. B} {\bf 682} (2009) 185. the PICASSO
  collaboration, arXiv:0907.0307 [astro-ex].

\bibitem{PICASSO11}
S.~Archambault et~al. {\em New J. Phys.} {\bf 13} (2011) 043006.
  arXiv:1011.4553 (physics.ins-det).

\bibitem{Chen}
T. P. Cheng, {\it Phys. Rev. D} {\bf 38}, 2869 (1988); H-Y. Cheng, {\it Phys.
  Lett. B} {\bf 219}, 347 (1989).

\bibitem{Dree00}
A. Djouadi and M. K. Drees, {\it Phys. Lett. B} {\bf 484}, 183 (2000); S.
  Dawson, {\it Nucl. Phys. B} {\bf 359}, 283 (1991); M. Spira {it et al}, {\it
  Nucl. Phys.} {\bf B453}, 17 (1995).

\bibitem{EFO00}
J.~Ellis, A.~Ferstl, and K.~A. Olive {\em Phys. Lett. B} {\bf 481} (2000) 304.

\bibitem{JDV06}
J. D. Vergados, On The Direct Detection of Dark Matter- Exploring all the
  signatures of the neutralino-nucleus interaction, hep-ph/0601064.

\bibitem{GTY09}
J.~Giedt, A.~Thomas, and R.~D. Young {\em Phys. Rev. Lett.} {\bf 103} (2009)
  201802.

\bibitem{MVE05}
C.~C. Moustakidis, J.~D. Vergados, and H.~Ejiri {\em Nucl. Phys. B} {\bf 727}
  (2005) 406. hep-ph/0507123.

\bibitem{XENON14}
E.~Aprile et~al. {\em J. Phys. G: Nucl. Part. Phys.} {\bf 41} (2014) 035201.
  [XENON100 Collaboration]; arXiv:1311.1088 (astro-ph.IM).

\bibitem{Li:forthcoming}
C.~Li et~al., {\it {to appear}}, .

\end{thebibliography}\endgroup

\end{document}